\PassOptionsToPackage{colorlinks=true,allcolors=blue}{hyperref}
\documentclass[%
  aps,%
  prapplied,%
  reprint,%
  superscriptaddress,%
  longbibliography%
]{revtex4-2}

\usepackage{graphicx}
\usepackage{amsmath,amssymb,bm}
\usepackage{microtype}
\usepackage{orcidlink}
\usepackage{siunitx}
\DeclareSIUnit\sample{Sa}
\usepackage{xcolor}

\newcommand{\REV}[1]{#1}

\newcommand{\figplaceholder}[2]{\fbox{\parbox[c][#1][c]{0.92\linewidth}{\centering\textcolor{red}{\texttt{\detokenize{#2}}}}}}
\newcommand{\fig}[2][\columnwidth]{\IfFileExists{#2}{\includegraphics[width=#1]{#2}}{\figplaceholder{4.5cm}{#2}}}

\begin{document}
\raggedbottom

\title{Absolute displacement calibration of a levitated nanoparticle with real-time heterodyne demodulation}

\author{Daniel Tandeitnik\,\orcidlink{0000-0003-3276-9335}}
\email{tandeitnik@gmail.com}
\affiliation{Department of Physics, Pontifical Catholic University of Rio de Janeiro, Rio de Janeiro 22451-900, Brazil}

\author{Gabriel Dias\,\orcidlink{0009-0001-6935-5585}}
\email{gabriel33828@gmail.com}
\affiliation{Department of Physics, Pontifical Catholic University of Rio de Janeiro, Rio de Janeiro 22451-900, Brazil}

\author{Thiago Guerreiro\,\orcidlink{0000-0001-5055-8481}}
\email{thguerreiro@gmail.com}
\affiliation{Department of Physics, Pontifical Catholic University of Rio de Janeiro, Rio de Janeiro 22451-900, Brazil}

\date{\today}

\begin{abstract}
We report a heterodyne scheme for the position readout of an optomechanical system, in particular an optically levitated nanoparticle, based on digital in-phase and quadrature demodulation on a field-programmable gate array. Compared to the standard homodyne approach, it offers three key advantages: it requires no phase lock, it produces a signal linearly proportional to the particle displacement, and its calibration is immune to drifts in the optical power of the local oscillator and of the scattered field. Combined with a vectorial model of the focal field, it measures the displacement in length units without free parameters. Through simultaneous homodyne and heterodyne measurements on the same trapped particle we demonstrate these advantages: the heterodyne follows a driven particle linearly up to amplitudes at which the homodyne response folds back, and calibrates the homodyne, whose standard thermometric calibration we find biased by \SI{33}{\percent}. Inverting equipartition, the calibrated readout measures the mass of the particle with an uncertainty below \SI{1}{\percent}, after removing a bias of up to \SI{20}{\percent} caused by absorption heating at \SI{1550}{nm}. The method can be used in any optomechanical system based on phase readout.
\end{abstract}

\maketitle

\section{Introduction}\label{sec:intro}

Optomechanical systems transduce mechanical motion into optical signals, and measurements based on these platforms, such as force, displacement, or acceleration sensing, \REV{rely on} the accuracy and linearity of this transduction process~\cite{braginsky1995quantum}. This operating principle is central to gravitational-wave interferometers~\cite{aasi2015advancedligo,abbott2016observation}, micromechanical sensors~\cite{aspelmeyer2014cavity}, optically levitated particles~\cite{magrini2021real,tebbenjohanns2021quantum,iwasaki2019electric,kremer2024all}, and nanomechanical oscillators operating in the quantum regime~\cite{thompson2008strong,chan2011laser,gavartin2012hybrid,purdy2013observation}. \REV{The sensitivity of many of these experiments is limited by quantum backaction or thermal noise, but the transduction sets how accurately the measured signal is converted into displacement and force.} In these systems, the mechanical position is encoded predominantly in the optical phase of the field interacting with the mechanical degree of freedom~\cite{clerk2010introduction,tebbenjohanns2019optimal}, and interferometric phase readout maps this phase modulation onto a measurable output signal. Beyond raw sensitivity, the practical utility of the readout is determined by the extent to which the measured signal can be quantitatively related to physical displacement, i.e., by the quality and robustness of its calibration.

In levitated optomechanics, the standard readout is balanced homodyne detection, wherein the scattered field (SF) is interferometrically combined with a phase-locked local oscillator (LO)~\cite{tebbenjohanns2019optimal,magrini2021real,tebbenjohanns2021quantum,militaru2022ponderomotive,maurer2023quantum}. The detector output is proportional to the sine of the optical phase, multiplied by a proportionality factor determined by the optical powers, detector gain, and spatial/temporal mode overlap, which are generally not known \emph{a priori}. Consequently, the transduction coefficient must be estimated \emph{in situ}. Commonly used protocols rely on external inputs: thermometric calibration presumes knowledge of the particle mass and the bath temperature~\cite{hebestreit2018calibration}, whereas calibration with an applied external force presumes a model of that force, such as the Coulomb force derived from the particle charge and a simulated electric field~\cite{ricci2019accurate}. Both approaches incur systematic bias when the phase excursion is not in the small-signal regime, because the intrinsic sinusoidal response redistributes spectral power into higher harmonics. Moreover, both procedures must be repeated whenever the optical powers are modified. An absolute calibration can be achieved in the quantum regime by exploiting motional sideband asymmetry~\cite{magrini2021real,tebbenjohanns2021quantum}, but this method is applicable only when the mechanical oscillator is prepared sufficiently close to its motional ground state.

We demonstrate that heterodyne detection implemented with real-time digital in-phase and quadrature (IQ) demodulation on a field-programmable gate array (FPGA) enables an absolute displacement readout that remains valid over arbitrary ambient pressure and displacement excursion. Because the optical phase is reconstructed via a four-quadrant arctangent applied to the demodulated quadratures, the measurement directly yields the optical phase in radians. The remaining mapping from phase to displacement is uniquely determined by the geometry of the trapping and collection fields,
\begin{subequations}\label{eq:absCalIntro}
    \begin{align}
        \Delta\phi &= 2k_{\mathrm{eff}}\,z,\\
        k_{\mathrm{eff}} &= \kappa k,
    \end{align}
\end{subequations}
where $k=2\pi/\lambda$ and the dimensionless geometric factor $\kappa\le 1$ is obtained from a vectorial model of the focal and collection fields~\cite{tandeitnik2026vectorial}. Notably, Eq.~\eqref{eq:absCalIntro} contains no dependence on the particle mass, temperature, pressure, or actuation parameters. The resulting readout is linear, up to a distortion by parasitic reflections that we measure and correct, and, furthermore, is independent of the optical power levels.

Heterodyne detection carries a factor-of-two penalty in imprecision and a limited dynamic range, which exclude it from the feedback loop in ground-state cooling experiments~\cite{magrini2021real,tebbenjohanns2021quantum}. We therefore use it as an out-of-loop reference. Operated on the same scattered field as the in-loop homodyne, it calibrates the homodyne transduction, the electrode force and the measurement rates of both detectors. It also shows that the thermometric calibration of the homodyne is biased low by the sinusoidal response of the detector, by tens of percent at the pressures where such calibrations are usually performed. By inverting equipartition it also measures the mass of the particle, which at \SI{1550}{nm} reveals a heating of the particle by absorption.

Heterodyne detection is already used in many levitated-particle experiments, for sideband thermometry~\cite{magrini2021real,tebbenjohanns2021quantum}, displacement sensing~\cite{rider2018} and the detection of squeezed light~\cite{magrini2022squeezed}. Our contribution is a real-time phase readout that is absolutely calibrated, \REV{without the plane-wave factor $4\pi/\lambda$~\cite{rider2018} or a known electric force~\cite{rider2018,tseng2025search},} and its use as a calibration reference for homodyne detection. \REV{Its linear range is useful wherever the motion is large, for example in force sensing with particles released from the trap~\cite{hebestreit2018staticforces}.}

\REV{The appendices give, in order, the vectorial model of $\kappa$ and the effective numerical aperture, the FPGA demodulation and the test of its phase scale, the noise budgets, the parasitic fields, the nonlinear bias of the thermometric calibration and the absorption heating of the particle.}

\section{Phase readout}\label{sec:readout}

\subsection{Homodyne detection}\label{sec:homodyne}

In balanced homodyne detection, the SF is interfered with a strong LO on a 50:50 beam splitter, and the difference photocurrent is recorded~\cite{bachor2019guide,loudon2000quantum}. With the two fields written as
\begin{subequations}\label{eq:fields}
    \begin{align}
        E_{\mathrm{SF}} &= \mathcal{E}_{\mathrm{SF}}\cos(\omega t+\phi_{\mathrm{SF}}),\\
        E_{\mathrm{LO}} &= \mathcal{E}_{\mathrm{LO}}\cos(\omega t+\phi_{\mathrm{LO}}),
    \end{align}
\end{subequations}
and co-polarized, the resulting homodyne output can be expressed as
\begin{equation}\label{eq:homodyneSignal}
    y(t) = V_0 \sin\!\left(\phi_{\mathrm{SF}} - \phi_{\mathrm{LO}}\right),
\end{equation}
where the fringe amplitude \(V_0\propto\mathcal{E}_{\mathrm{SF}}\mathcal{E}_{\mathrm{LO}}\) also depends on the gain and responsivity of the detector. \REV{Here and below, the detector output is the photocurrent averaged over the response time of the detector.} An axial displacement \(z\) shifts the phase of the back-scattered field by \(2k_{\mathrm{eff}}z\), and the LO phase is actively locked so that \(\phi_{\mathrm{SF}}-\phi_{\mathrm{LO}} = 2k_{\mathrm{eff}}z+\phi_0\), with \(\phi_0\) a small residual lock offset. In the small-signal regime, the displacement-to-voltage transduction coefficient therefore reads
\begin{equation}\label{eq:Chom}
    C_{\mathrm{hom}} = 2k_{\mathrm{eff}}V_0|\cos\phi_0| \propto \mathcal{E}_{\mathrm{SF}}\mathcal{E}_{\mathrm{LO}}.
\end{equation}
Notably, \(C_{\mathrm{hom}}\) depends multiplicatively on the optical amplitudes of both fields and is thus sensitive to slow drifts in either \(\mathcal{E}_{\mathrm{LO}}\) or \(\mathcal{E}_{\mathrm{SF}}\). Consequently, it cannot, in general, be determined reliably from nominal parameters alone and typically necessitates \emph{in situ} calibration.

Beyond the linear regime, the inherently sinusoidal nature of the phase-to-voltage transduction generates a systematic calibration bias. For a coherently driven displacement \(z=A\sin(\omega_{\mathrm{d}}t)\), application of the Jacobi--Anger expansion~\cite{arfken2013mathematical} gives
\begin{equation}\label{eq:jacobiAnger}
    y(t) = 2V_0\sum_{n=0}^{\infty} J_{2n+1}(\phi_A)\sin\!\big((2n+1)\omega_{\mathrm{d}}t\big),
\end{equation}
where \(\phi_A = 2k_{\mathrm{eff}}A\) and $J_n$ are Bessel functions of the first kind. It follows that the spectral component at the drive frequency is proportional to \(2V_0J_1(\phi_A)\), rather than the linear-response expression \(V_0\phi_A\). As a result, the fundamental response exhibits saturation: it reaches a maximum near \(\phi_A \simeq 1.84\) and decreases for larger excursions, with the associated signal power transferred into higher-order odd harmonics.

For thermally driven motion, the phase is well approximated as Gaussian with variance \(\sigma_\phi^2\). In this case, Bussgang's theorem~\cite{bussgang1952crosscorrelation} decomposes the output into a term linearly proportional to the phase and a distortion term that is uncorrelated with the phase (Appendix~\ref{app:bussgang}). A thermometric calibration fits a Lorentzian to the homodyne power spectral density (PSD)~\cite{gieseler2021optical,hebestreit2018calibration} and, with the mass and the temperature taken as known, infers from its area a coefficient \(C_{\mathrm{hom}}^{\mathrm{th}}\). Because the fit is sensitive only to the linear contribution, \(C_{\mathrm{hom}}^{\mathrm{th}}\) underestimates the small-signal coefficient \(C_{\mathrm{hom}}\) of Eq.~\eqref{eq:Chom} by the factor
\begin{equation}\label{eq:bussgang}
    \frac{C_{\mathrm{hom}}^{\mathrm{th}}}{C_{\mathrm{hom}}} = e^{-\sigma_\phi^2/2}.
\end{equation}
This constitutes the thermal analogue of \(2J_1(\phi_A)/\phi_A\). Without feedback the thermal phase variance is set by the mass, the axial angular frequency $\Omega$ of the trap and the temperature, not by the pressure. For a \SI{156}{nm} silica particle at \(\Omega/2\pi\approx\SI{45}{\kilo\hertz}\) the bias is about \SI{30}{\percent}.

\subsection{Heterodyne detection with IQ demodulation}\label{sec:heterodyne}

In heterodyne detection the LO is frequency-shifted by $\Delta\omega$, and the balanced detector outputs a beat note at this frequency whose phase encodes $\phi_{\mathrm{SF}}$. Demodulation is implemented by mixing the detector output with $\cos(\Delta\omega t)$ and $-\sin(\Delta\omega t)$ followed by suppression of the image term at $2\Delta\omega$~\cite{meade1983lockin,lyons2011understanding}, yielding the in-phase and quadrature components
\begin{subequations}\label{eq:IQ}
    \begin{align}
        I &\propto \mathcal{E}_{\mathrm{LO}}\mathcal{E}_{\mathrm{SF}}\sin\psi, \\ Q &\propto \mathcal{E}_{\mathrm{LO}}\mathcal{E}_{\mathrm{SF}}\cos\psi,
    \end{align}
\end{subequations}
where $\psi=\phi_{\mathrm{SF}}-\phi_{\mathrm{LO}}$. The measured phase is then obtained via the four-quadrant arctangent,
\begin{equation}\label{eq:demodPhase}
    \phi_{\mathrm{meas}} = \arg(Q + iI) = \phi_{\mathrm{SF}} - \phi_{\mathrm{LO}}.
\end{equation}
Because the common amplitude factor $\mathcal{E}_{\mathrm{LO}}\mathcal{E}_{\mathrm{SF}}$ cancels in Eq.~\eqref{eq:demodPhase}, the phase transduction is, to first order, independent of the LO and SF optical powers. Consequently, after phase unwrapping, $\phi_{\mathrm{SF}}$ is recovered linearly over arbitrary excursions. The LO phase, as well as any residual mismatch between the digital reference and the actual beat frequency, contributes only slowly varying additive terms, which are removed after unwrapping. The noise of the readout follows from the same amplitude: additive noise of spectral density $S_n$ on each quadrature becomes a phase noise $S_n/|Q+iI|^2$, set by the collected power and, unlike the homodyne sensitivity, independent of the operating point. We use this relation in Sec.~\ref{sec:noise} to express the measured noise floors in phase units.

\subsection{Absolute displacement calibration}\label{sec:absolute}

The FPGA outputs the phase in radians with an exactly known scaling, which we validate end-to-end against homodyne interference fringes (Appendix~\ref{app:fpga}). Converting it into displacement requires the factor $\kappa$ of Eq.~\eqref{eq:absCalIntro}, which we compute with a vectorial model of the trapping and collection fields~\cite{tandeitnik2026levitationtoolbox}. It depends on the numerical aperture and the filling factor. For the numerical aperture we use the effective value of the focus, determined \emph{in situ} from the ratios of the three mechanical frequencies (Appendix~\ref{app:kappa}). For our trap, $\kappa = \num{0.8582(13)}$, i.e.\ $2k_{\mathrm{eff}} = \SI{6.958(10)e6}{rad/m}$, a relative uncertainty of \SI{0.15}{\percent}.

Under an absolute phase readout, parameters that ordinarily enter as calibration inputs instead become measurement outputs. In particular, the equipartition relation, \(m\Omega^2\langle z^2\rangle = k_BT\), with $T$ the gas temperature, assumed here to be also that of the centre-of-mass motion (see Sec.~\ref{sec:mass}), and $k_B$ the Boltzmann constant, yields the oscillator mass,
\begin{equation}\label{eq:mass}
    m = \frac{k_BT\,(2k_{\mathrm{eff}})^2}{\Omega^2\langle\phi^2\rangle},
\end{equation}
where \(\langle\phi^2\rangle\) and \(\Omega\) are extracted from the phase PSD. The same records also characterize the detection itself, through its measurement rate, which quantifies how fast the detector acquires information on the position. It is given by~\cite{clerk2010introduction}
\begin{subequations}\label{eq:GmeasDef}
    \begin{align}
        \Gamma_{\mathrm{meas}} &= \frac{z_{\mathrm{zpf}}^2}{2S_{zz}},\\
        z_{\mathrm{zpf}} &= \sqrt{\frac{\hbar}{2m\Omega}},
    \end{align}
\end{subequations}
where $z_{\mathrm{zpf}}$ is the zero-point fluctuation amplitude and $S_{zz}$ the one-sided imprecision. In terms of the phase it reads
\begin{equation}\label{eq:Gmeas}
    \Gamma_{\mathrm{meas}} = \frac{\hbar\Omega\,\langle\phi^2\rangle}{4k_BT\,S_{\phi\phi}^{\mathrm{imp}}},
\end{equation}
which is independent of both the mass and \(k_{\mathrm{eff}}\). Consequently, \(\Gamma_{\mathrm{meas}}\) is determined solely by the ratio between the thermally induced phase variance and the white phase-noise floor \(S_{\phi\phi}^{\mathrm{imp}}\). This formulation is therefore applicable to any detector with a known phase scale, including a homodyne detector once its phase response has been calibrated against the heterodyne readout.

\section{Experimental setup}\label{sec:setup}

\begin{figure}[tbp]
    \centering
    \fig{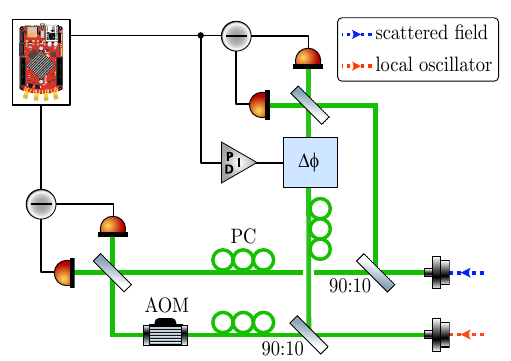}
    \caption{Experimental configuration of the detection. The SF and the LO are split by 90:10 fibre couplers between the homodyne and the heterodyne branch, and both balanced detectors are digitized by the same demodulation board. $\Delta\phi$, fibre stretcher; PID, proportional-integral-derivative controller; AOM, acousto-optic modulator; PC, polarization controllers.}
    \label{fig:setup}
\end{figure}

The detection subsystem is depicted in Fig.~\ref{fig:setup}. A $\mathrm{SiO_2}$ particle with a nominal diameter of \SI{156(5)}{\nano\meter} is trapped at \SI{1550}{\nano\meter} by an aspheric objective of effective focal length \SI{0.53(1)}{\milli\meter} and nominal numerical aperture 0.75, illuminated by a linearly polarized beam of radius \SI{0.47}{\milli\meter}. The effective numerical aperture of the focus and the filling factor are determined \emph{in situ} (Sec.~\ref{sec:absolute}). The back-scattered field is collected confocally~\cite{magrini2021real} and coupled into a single-mode fibre, whose fundamental mode defines the measurement mode entering $\kappa$. The LO is derived from the trapping beam. Two 90:10 fibre couplers send \SI{90}{\percent} of the SF to the homodyne branch and \SI{90}{\percent} of the LO to the heterodyne branch, whose sensitivity at low LO power is limited by detector electronic noise. In the heterodyne branch an acousto-optic modulator shifts the LO by $\Delta\omega/2\pi=\SI{10}{\mega\hertz}$. A knife-edge detector in the forward direction records the three modes of motion.

Two Red Pitaya boards run the electronics. The first locks the homodyne LO phase with a PID controller driving the fibre stretcher. The second, the demodulation board, digitizes the heterodyne and homodyne signals simultaneously and demodulates the heterodyne in real time. The particle is driven through two electrodes, the metallic housing of the trapping lens and a ring electrode mounted between the trapping and collection lenses. The DC monitor output of the homodyne detector, proportional to the LO power, is recorded with an oscilloscope.

\section{Absolute calibration in practice}\label{sec:results}

We first verify the heterodyne readout against the folds of the homodyne (Sec.~\ref{sec:linearity}) and under a strong coherent drive, which also calibrates the electrode force (Sec.~\ref{sec:driven}). We then use it as a reference: it calibrates the homodyne (Sec.~\ref{sec:transfer}), measures the mass of the particle and its heating by absorption (Sec.~\ref{sec:mass}), and gives the measurement rate and noise budget of both detectors (Sec.~\ref{sec:noise}). Unless stated otherwise the particle is at about \SI{10}{\milli\bar} with feedback off, and each data set consists of simultaneous records of the heterodyne phase and the homodyne signal.

\subsection{Linearity and a theory-free check of the phase scale}\label{sec:linearity}

\begin{figure}[t]
    \centering
    \fig{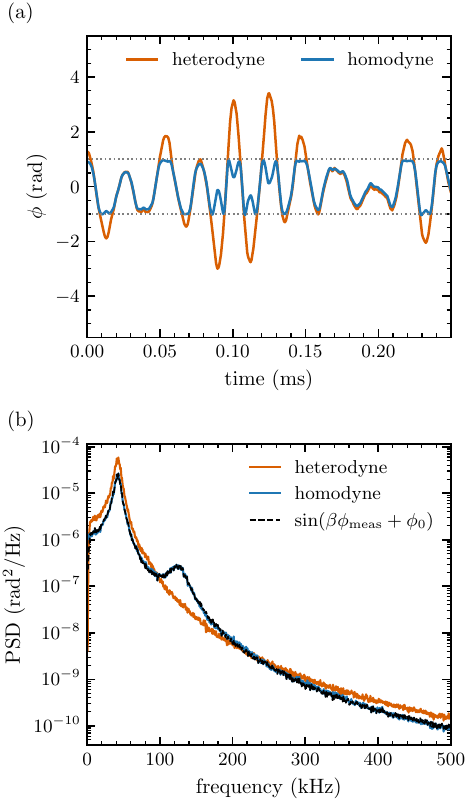}
    \caption{Simultaneous heterodyne and homodyne records of the same thermally driven particle. (a)~Time traces: heterodyne phase (orange) and homodyne signal normalized by its fringe amplitude $V_0$ (blue); the homodyne folds whenever the phase exceeds $\pm\pi/2$. (b)~PSDs averaged over thirty records, in phase units. The homodyne shows odd harmonics of the axial frequency; the dashed curve is the homodyne PSD computed from the heterodyne record through Eq.~\eqref{eq:transfer}, with no further parameters.}
    \label{fig:linearity}
\end{figure}

We first test that the heterodyne returns the optical phase linearly and with the correct scale, by comparing it with the homodyne recorded at the same time. Figure~\ref{fig:linearity}(a) shows simultaneous time traces for a particle of axial frequency $\Omega/2\pi \approx \SI{43}{\kilo\hertz}$. The thermal phase excursion reaches several radians; the heterodyne follows it linearly, whereas the homodyne is bounded by $\pm V_0$ and folds back whenever $|\phi|$ exceeds $\pi/2$. In the spectrum [Fig.~\ref{fig:linearity}(b)] the folding appears as odd harmonics of the axial frequency, absent from the heterodyne.

Because both branches observe the same optical phase, the homodyne signal must be the sine of the heterodyne phase. We fit, in segments of \SI{2}{ms} to allow for slow wander of the lock point,
\begin{equation}\label{eq:transfer}
    y(t) = V_0\sin\!\big(\beta\,\phi_{\mathrm{meas}}(t) + \phi_0\big) + y_0,
\end{equation}
where $V_0$ is the homodyne fringe amplitude, $\phi_0$ the lock offset, $y_0$ an electronic offset, and $\beta$ a scale factor that must equal unity. The homodyne folds at exactly $\pm\pi/2$ of the true optical phase, so its fold points act as an absolute ruler for the heterodyne phase, independent of $\kappa$ and of any model. The dashed curve in Fig.~\ref{fig:linearity}(b), computed from the heterodyne record through Eq.~\eqref{eq:transfer}, reproduces the homodyne harmonics. Across runs and sessions $\beta$ ranged from 0.98 to 1.16, which is caused by a parasitic back-reflection with an amplitude about four times smaller than the SF (Appendix~\ref{app:parasitic}). The records used here contain the phase but not the full I/Q vector, so the distortion cannot be removed by the circle fit used in Sec.~\ref{sec:mass}. Inverting the parasitic map instead, with the amplitude ratio measured \SI{20}{min} earlier, brings $\beta$ to unity within \SI{0.3}{\percent} and halves the fit residuals. The scale factor $\beta$ of each run corrects the slope of this map only to first order: the thermal variance of the $\beta$-corrected phase reads \SI{4}{\percent} low against the circle fit of the same particle. The mass (Sec.~\ref{sec:mass}) and the drive amplitude (Sec.~\ref{sec:driven}) therefore use the circle fit or the inverted map. Because the check requires excursions that reach the folds, it is available only without feedback.

\subsection{Driven response and absolute electrode calibration}\label{sec:driven}

A coherent drive tests the linearity of the heterodyne far beyond the folds of the homodyne. Because the heterodyne measures the displacement in length units, the same measurement calibrates the force that the electrodes exert on an electrically charged particle. A sinusoidal voltage applied to the electrodes at $\omega_{\mathrm{d}}$ drives a displacement of amplitude $A$, which the heterodyne measures as $A = \phi_A/2k_{\mathrm{eff}}$. To show the shortcoming of a homodyne readout for force calibration, we drive the particle slightly above its axial resonance with ten increasing amplitudes, and record four pairs of heterodyne and homodyne traces at each amplitude. Lock-in detection at $\omega_{\mathrm{d}}$ and its harmonics gives $\phi_A$ and the homodyne harmonic amplitudes $Y_n$.

The heterodyne phase is corrected for the parasitic field with Eq.~\eqref{eq:phasePF}, with the amplitude ratio $r$ of the scattered and parasitic fields from a circle fit taken before the series and the slowly drifting phase of the parasite followed every \SI{2}{ms} with the homodyne folds of the same records. A single scale factor $\beta$ would not suffice here, because the distortion depends on the drive amplitude: the fold fit gives $\beta = 1.07$ at the smallest amplitude and 1.01 at the largest, and 1.000 to 1.003 at all amplitudes with the lock held after the correction.

Since the thermal motion is present as well, the coherent part of the homodyne output, averaged over the Gaussian thermal phase, is
\begin{figure}[t]
    \centering
    \fig{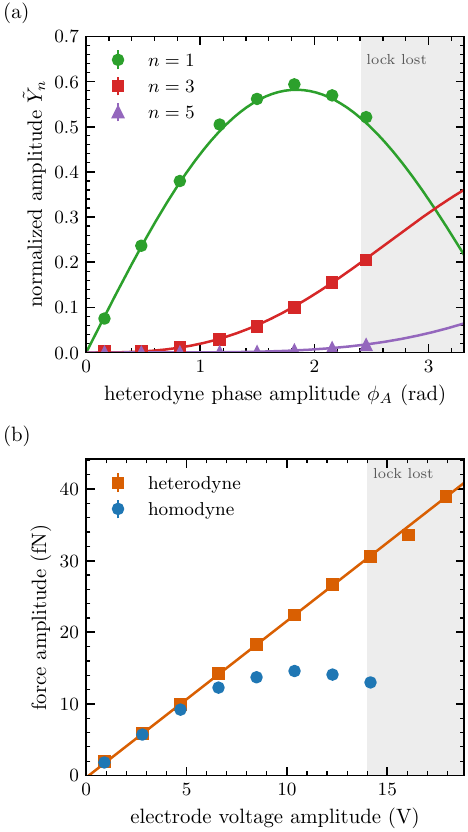}
    \caption{Electrode drive at $\approx\SI{48}{\kilo\hertz}$, recorded in both branches. (a)~Homodyne harmonics $n = 1, 3, 5$, normalized by the prefactor of Eq.~\eqref{eq:Yn}, against the heterodyne phase amplitude $\phi_A$. Lines: $|J_n(\phi_A)|$, without free parameters. Shaded: beyond the first zero of $J_0$, where the lock signal changes sign. Points with the lock lost are not shown. (b)~Force amplitude against electrode voltage, from the heterodyne (orange, linear fit) and from the homodyne with its thermometric calibration (blue).}
    \label{fig:driven}
\end{figure}
\begin{equation}\label{eq:coherent}
    \langle y(t)\rangle = V_0\,e^{-\sigma_\phi^2/2}\sin\!\big(\phi_A\sin\omega_{\mathrm{d}}t+\phi_0\big),
\end{equation}
and its odd harmonics are [Eq.~\eqref{eq:jacobiAngerOffset}]
\begin{equation}\label{eq:Yn}
    Y_n = 2V_0|\cos\phi_0|\,e^{-\sigma_\phi^2/2}\,|J_n(\phi_A)|,
\end{equation}
with $V_0$, $\phi_0$ and $\sigma_\phi$ measured in the same records. Figure~\ref{fig:driven}(a) compares the measured harmonics with this prediction, which has no free parameter. The fundamental follows $J_1$ through its maximum at \SI{1.84}{rad} to within \SI{2.6}{\percent} rms, and the third harmonic follows $|J_3|$ within \SI{4}{\percent}. Beyond the first zero of $J_0$, at \SI{2.4}{rad}, the signal on which the phase lock acts changes sign. The lock still held at \SI{2.45}{rad} but was lost at the two largest amplitudes, whose homodyne points are not shown. The heterodyne, which needs no lock, is unaffected [Fig.~\ref{fig:driven}(b)].

The heterodyne also calibrates the actuation without a model of the force, that is, without knowing the charge of the particle or the field of the electrodes. With the mass and the mechanical susceptibility from the fitted $\Omega$ and gas damping rate $\gamma$ of the same records, the force amplitude follows as~\cite{gieseler2021optical}
\begin{subequations}\label{eq:force}
    \begin{align}
        F &= m\sqrt{(\Omega_A^2-\omega_{\mathrm{d}}^2)^2 + \gamma^2\omega_{\mathrm{d}}^2}\;A ,\\
        \Omega_A^2 &= \Omega^2\Big(1+\frac{3}{4}\epsilon A^2\Big),
    \end{align}
\end{subequations}
where $\Omega_A$ includes the softening of the axial restoring force $m\Omega^2 z(1+\epsilon z^2)$, with $\epsilon = \SI{-1.7e11}{m^{-2}}$ computed with the vectorial model~\cite{tandeitnik2026levitationtoolbox} (the quadratic term of the force changes it by less than \SI{1}{\percent}). It raises the force by \SI{1.6}{\percent} at $A = \SI{250}{nm}$ and by \SI{5}{\percent} at \SI{440}{nm}. With both corrections the forces are linear in the drive voltage within \SI{1}{\percent} up to \SI{440}{nm}, except at \SI{390}{nm}, where the lock point was slipping during the records and the fold fit could not follow it, and the slope gives the force per volt, $C_{\mathrm{NV}} = \SI{2.19(3)e-15}{N/V}$. The mass is the inertial mass of the particle, from a power series measured just before (Sec.~\ref{sec:mass}), since the equipartition mass of the driven records would be biased low by the heating of the particle~\cite{millen2014nanoscale}. For comparison, Fig.~\ref{fig:driven}(b) shows the force that a homodyne readout alone would report, using its fundamental and its thermometric calibration. It equals the true force times $2J_1(\phi_A)/\phi_A$: it saturates and then decreases, so that at the largest amplitude with the lock held it reports only \SI{43}{\percent} of the true force. Being non-monotonic, this response cannot be inverted. An electrode calibration based on the homodyne signal, such as that of Ref.~\cite{ricci2019accurate}, therefore requires drive amplitudes well inside the linear range of the homodyne, whereas the heterodyne readout has no such restriction.

\subsection{Calibration transfer to the homodyne}\label{sec:transfer}

\begin{figure}[tbp]
    \centering
    \fig{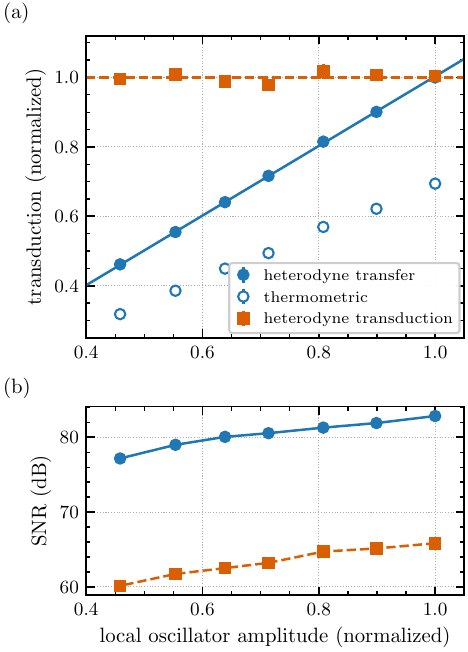}
    \caption{Transduction and signal-to-noise ratio as a function of the LO amplitude normalized to its maximum, that is, the square root of the LO power relative to full power, measured simultaneously in both branches. (a)~Homodyne coefficient from the heterodyne transfer (filled circles) with a proportional fit, and from a thermometric fit (open circles), both normalized to the former at full LO. The heterodyne transduction is normalized to its mean (squares). (b)~Thermal peak height over the white floor, with lines joining the points.}
    \label{fig:loScan}
\end{figure}

With its phase scale verified, the heterodyne calibrates the homodyne, that is, it gives the homodyne voltage per unit displacement without an assumed mass or temperature. The transfer fit of Sec.~\ref{sec:linearity} gives the homodyne voltage per radian of true optical phase, $V_0$, and with Eq.~\eqref{eq:Chom} the homodyne coefficient is $C_{\mathrm{hom}} = \SI{8.37(2)}{mV/nm}$, with an accuracy set by $\kappa$.

A standard thermometric fit to the homodyne PSD with the nominal mass returns \SI{5.87(28)}{mV/nm}, \SI{30}{\percent} below $C_{\mathrm{hom}}$. This is dominated by the \SI{33}{\percent} nonlinear bias of Eq.~\eqref{eq:bussgang}: with the equipartition mass of the particle at the same trap power, measured with the circle fit (Sec.~\ref{sec:mass}), the thermometric value is $0.665(3)$ of $C_{\mathrm{hom}}$, in agreement with the predicted $e^{-\sigma_\phi^2/2} = 0.666$ for the measured $\sigma_\phi = \SI{0.90}{rad}$, with no free parameters. The rest is the difference between the measured and nominal masses.

Figure~\ref{fig:loScan} shows both calibrations as the LO power is reduced in steps from \SI{100}{\percent} to \SI{21}{\percent}, the power being read from the DC monitor output of the detector. The homodyne coefficient scales with the LO amplitude, as Eq.~\eqref{eq:Chom} requires, and the thermometric values track it at the constant nonlinear bias. This proportionality lets the calibration follow changes of the LO power through the detector monitor. The heterodyne transduction remains constant within \SI{1.6}{\percent}. Both signal-to-noise ratios increase with the LO [Fig.~\ref{fig:loScan}(b)], from 77 to \SI{84}{\decibel} in homodyne and from 61 to \SI{67}{\decibel} in heterodyne, because both detectors are partly limited by electronic noise at these powers (Sec.~\ref{sec:noise}).

\subsection{Particle mass and absorption heating}\label{sec:mass}

\begin{figure*}[t]
    \centering
    \fig[\textwidth]{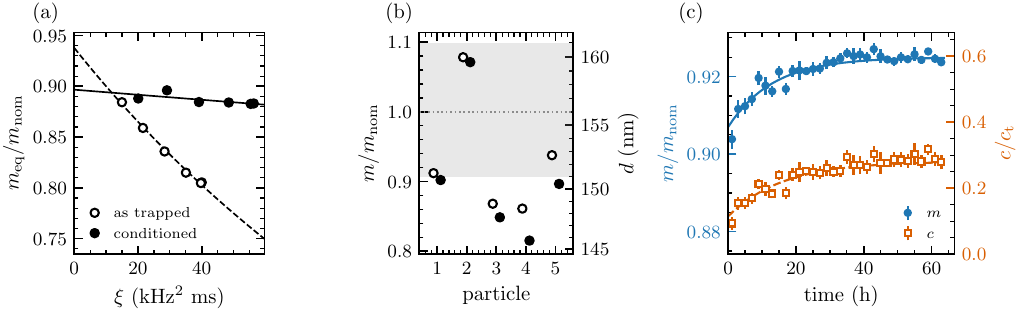}
    \caption{Masses of particles of nominal diameter \SI{156(5)}{nm}, relative to the nominal mass $m_{\mathrm{nom}}$ (density \SI{1850}{kg/m^3}). (a)~Equipartition mass of particle 5 against the heating parameter $\xi$, as trapped (open circles) and after conditioning (filled circles), with fits of Eq.~\eqref{eq:massSeries} drawn to $\xi=0$, where they give $m$. (b)~Mass of the five particles, as trapped and after conditioning (symbols as in (a)). The band marks the nominal size distribution, and the right axis gives the diameter at the nominal density. (c)~Mass and heating coefficient of particle 5 after conditioning, left in laboratory air, as \SI{2}{h} means with saturating-exponential fits. The heating coefficient is relative to its value as trapped, $c_{\mathrm{t}}$.}
    \label{fig:massCampaign}
\end{figure*}

\begin{table*}[t]
    \caption{Mass $m$ and heating coefficient $c$ of five particles of nominal diameter \SI{156(5)}{nm}, as trapped and after conditioning, from the extrapolation of Fig.~\ref{fig:massCampaign}; $d$ is the diameter at the nominal density, $m_{\mathrm{nom}} = \SI{3.677e-18}{kg}$. Uncertainties are statistical (fit of each series), and $\kappa$ and the temperature add \SI{0.3}{\percent} each.}
    \label{tab:mass}
    \begin{ruledtabular}
    \begin{tabular}{ccccccc}
        & \multicolumn{2}{c}{as trapped} & \multicolumn{3}{c}{conditioned} & \\
        particle & $m/m_{\mathrm{nom}}$ & $c$\footnotemark[1] & $m/m_{\mathrm{nom}}$ & $m$ ($10^{-18}$\,kg) & $c$\footnotemark[1] & $d$ (nm) \\
        \hline
        1 & 0.912(3) & 3.22(11) & 0.902(3) & 3.318(11) & 0.68(15) & 150.7 \\
        2 & 1.078(7) & 3.64(22) & 1.071(6) & 3.940(22) & 0.77(14) & 159.6 \\
        3 & 0.868(6) & 4.03(23) & 0.849(6) & 3.121(22) & 0.72(21) & 147.7 \\
        4 & 0.861(6) & 5.92(24) & 0.815(5) & 2.998(19) & 0.25(21) & 145.7 \\
        5 & 0.938(4) & 4.20(16) & 0.897(5) & 3.298(20) & 0.28(14) & 150.4 \\
        \hline
        mean, s.d. & & & 0.907(99) & 3.33(36) & & 150.8(53) \\
    \end{tabular}
    \end{ruledtabular}
    \footnotetext[1]{In units of $10^{-3}\,(\mathrm{kHz^2\,ms})^{-1}$.}
\end{table*}

The absolute readout turns Eq.~\eqref{eq:mass} into a measurement of the mass of the particle. However, a trapped particle absorbs part of the trapping light, and gas molecules leaving its hot surface heat the centre of mass to $T_{\mathrm{cm}} > T$~\cite{millen2014nanoscale}, so that Eq.~\eqref{eq:mass} evaluated with $T$ returns the equipartition mass $m_{\mathrm{eq}} = m\,T/T_{\mathrm{cm}}$, with $m$ the mass of the particle. At \SI{1550}{nm} the effect is large: St\"ober silica contains water~\cite{ricci2022chemical}, which absorbs strongly at this wavelength~\cite{hale1973optical}, and the heating depends on the wavelength through the water absorption bands.

The absorbed power scales with the trapping power $P$ and the gas cooling with the pressure $p$~\cite{millen2014nanoscale}, so to first order (Appendix~\ref{app:heating})
\begin{subequations}\label{eq:heating}
    \begin{align}
        \frac{T_{\mathrm{cm}}}{T} &= 1 + c\,\xi,\\
        \xi &= \frac{f_z^2}{\gamma},
    \end{align}
\end{subequations}
with the axial frequency $f_z = \Omega/2\pi$, the heating coefficient $c$ and the heating parameter $\xi\propto P/p$, since $f_z^2\propto P$ and $\gamma\propto p$~\cite{gieseler2021optical}. Both $f_z$ and $\gamma$ are fitted in the same record as $\langle\phi^2\rangle$. We therefore measure the mass at several trapping powers at fixed pressure, from \SI{925}{mW} down to \SI{350}{mW} of laser output and back, and extrapolate to $\xi\to0$,
\begin{equation}\label{eq:massSeries}
    m_{\mathrm{eq}} = \frac{m}{1 + c\,\xi}.
\end{equation}
The repeated point detects any change of the particle during the series. Each point uses the phase obtained from a circle fit to the demodulated I/Q vector, which removes the parasitic distortion without $\beta$ [Fig.~\ref{fig:iqPlane}], and $\kappa$ is evaluated from a spectrum of the three modes taken just before the series.

Figure~\ref{fig:massCampaign}(a) shows such series for one particle, and Table~\ref{tab:mass} gives the results for five particles from the same batch. A freshly trapped particle, brought from ambient pressure to the working pressure of \SIrange{7}{9}{mbar}, heats strongly. At \SI{925}{mW} its centre of mass is \SIrange{35}{77}{K} above the gas, and Eq.~\eqref{eq:mass} evaluated with the gas temperature returns only \SIrange{79}{89}{\percent} of $m$. The absorber can be removed by conditioning the particle, that is, holding it for \SIrange{5}{10}{min} at \SI{1e-4}{mbar} and \SI{1.2}{W} of laser output, where gas cooling is negligible. Back at the working pressure, this lowers $c$ by a factor between 5 and 24, and the extrapolation to $\xi = 0$ shrinks to about \SI{1}{\percent}, small enough that its result no longer depends on the form of the heating law. We therefore quote the conditioned masses.

A conditioned particle left in laboratory air at about \SI{10}{mbar} slowly takes up the absorber again, most likely water. Particle 5, followed for \SI{64}{h} [Fig.~\ref{fig:massCampaign}(c)], regained mass and heating together, levelling off below its values as trapped. We therefore condition every particle shortly before its mass is measured, and fit $m$ and $c$ in every series.

The five conditioned masses scatter by \SI{11}{\percent} [Fig.~\ref{fig:massCampaign}(b)], which corresponds to diameters of \SI{151(5)}{nm} at the nominal density, the size spread specified by the manufacturer. Each mass is measured to better than \SI{1}{\percent}, so the measurement resolves the particle-to-particle distribution. The mean mass lies slightly below the nominal one, consistent with a slightly smaller mean diameter or a lower density. The damping of the three modes confirms that the particles are single spheres (Appendix~\ref{app:heating}).

Other mass measurements of levitated particles rely on a known force, the Coulomb force on a known charge in a simulated electrode field~\cite{ricci2019accurate}, or on a model of the trap anharmonicity combined with equipartition~\cite{zhang2026mass}. Ours needs neither a force nor a trap model, and it measures and removes the heating bias that equipartition methods, which assume the centre of mass at the gas temperature, do not account for.

\subsection{Measurement rate, imprecision and noise budget}\label{sec:noise}

\begin{figure}[tbp]
    \centering
    \fig{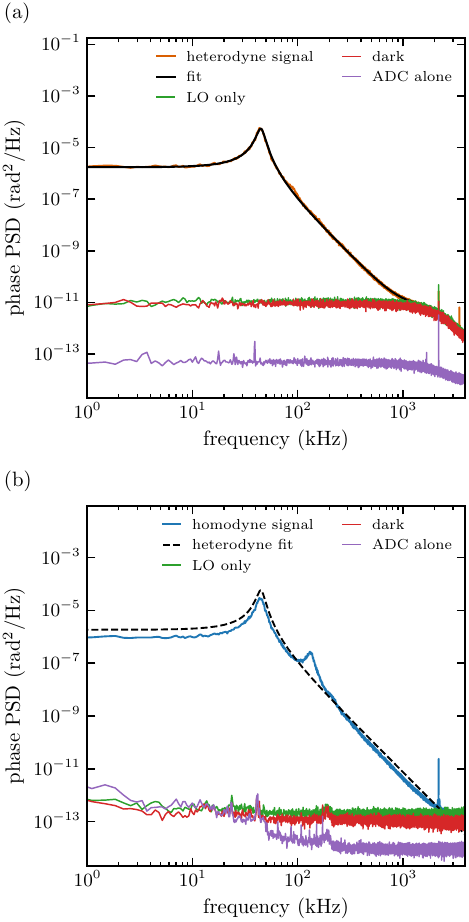}
    \caption{Phase spectra of (a) the heterodyne and (b) the homodyne with the particle trapped, with the SF blocked (LO only), with all light blocked (dark) and with the inputs of the analog-to-digital converter (ADC) terminated, all converted to phase units. The dashed line in (b) is the heterodyne fit of (a), the thermal spectrum used for both detectors.}
    \label{fig:noise}
\end{figure}

The absolute phase scale also characterizes the detectors. Equation~\eqref{eq:Gmeas} gives the measurement rate of each detector from the thermal phase variance and its white phase floor, with $T$ replaced by the centre-of-mass temperature $T_{\mathrm{cm}}$ known from the power series of the particle (Sec.~\ref{sec:mass}). The power series of this particle, which was not conditioned, was taken the next day: the ratio $m/m_{\mathrm{eq}}$ of these records gives $T_{\mathrm{cm}}/T = 1.08$, while the heating law of the series gives 1.15 at their operating point, and the difference, \SI{6}{\percent}, is taken as a systematic on the rates. For both detectors the thermal variance is taken from the fit of the heterodyne spectrum [Fig.~\ref{fig:noise}(a), dashed in (b)], since the homodyne spectrum alone underestimates it (Sec.~\ref{sec:transfer}). The same fit gives the heterodyne floor. The homodyne floor, buried under the thermal peak, is measured with the SF blocked and converted to phase with the fringe amplitude of Sec.~\ref{sec:linearity}. Since both detectors observe the same motion, the backaction cancels in the ratio of their efficiencies, which therefore equals the ratio of their phase floors, 55 for the run of Fig.~\ref{fig:noise}, independent of the mass, the temperature and the scattering model.

Blocking the SF, the LO or both separates each floor into its optical and electronic parts (Fig.~\ref{fig:noise}). In our setup both detectors are limited by electronic noise, the heterodyne more strongly, so that without it the homodyne efficiency would double (Appendix~\ref{app:noise}).

With the photon scattering rate $\Gamma_{\mathrm{S}}$ from the vectorial model~\cite{tandeitnik2026vectorial,tandeitnik2026levitationtoolbox}, scaled with the measured $\Omega$ and $m$, the recoil heating rate $\Gamma_{\mathrm{bk}} = C_z\hbar k^2\Gamma_{\mathrm{S}}/(2m\Omega)$, with $C_z$ the generalized geometric factor of Ref.~\cite{tandeitnik2026vectorial}, is $\Gamma_{\mathrm{bk}}/2\pi = \SI{11.1}{kHz}$ for the particle of Fig.~\ref{fig:noise}. It gives the detection efficiency $\eta = \Gamma_{\mathrm{meas}}/\Gamma_{\mathrm{bk}}$. We find $\eta = 0.08$ for the homodyne and $1.5\times10^{-3}$ for the heterodyne. Since $\Gamma_{\mathrm{bk}}$ comes from the scattering model, a comparison with reheating measurements with feedback switched off~\cite{jain2016direct,magrini2021real,tebbenjohanns2021quantum}, which would also reveal any heating beyond photon recoil, is left for future work.

\section{Conclusions}\label{sec:conclusions}

We have shown that real-time digital IQ demodulation of a heterodyne signal, combined with a vectorial model of the trapping and collection fields constrained \emph{in situ}, measures the displacement of a levitated nanoparticle in length units without free parameters, at any pressure and excursion. Alongside a homodyne detector, simultaneous records calibrate the homodyne, the electrode force and the mass of the particle, and give the efficiencies of both detectors. At \SI{1550}{nm} it also reveals that the particle is heated by absorption, a bias that equipartition-based methods must account for.

The accuracy is currently set by $\kappa$, through the effective numerical aperture, and by this heating. The absolute readout also opens further measurements: the density of the particle, an out-of-loop measurement of the occupation of feedback-cooled states, and the anharmonicity of the optical potential. Since the digital pipeline uses standard cores on a commercial FPGA platform, the scheme can be added to the heterodyne branch present in many levitated-optomechanics setups at the cost of a demodulation stage.

\section*{Acknowledgments}

We acknowledge Lukas Novotny and Joanna Zieli\'{n}ska for help with the optical setup.

We acknowledge support from the Coordena\c{c}\~ao de Aperfei\c{c}oamento de Pessoal de N\'ivel Superior - Brasil (CAPES) - Finance Code 001, Conselho Nacional de Desenvolvimento Cient\'ifico e Tecnol\'ogico (CNPq), Funda\c{c}\~ao de Amparo \`a Pesquisa do Estado do Rio de Janeiro (FAPERJ Scholarship No. E-26/203.727/2025), Funda\c{c}\~ao de Amparo \`a Pesquisa do Estado de S\~ao Paulo (FAPESP processo 2021/06736-5), the Serrapilheira Institute (grant No. Serra -- 2211-42299) and StoneLab.

\appendix
\twocolumngrid

\section{Geometric factor and effective numerical aperture}\label{app:kappa}

The factor $\kappa$ follows from the semiclassical vectorial scattering framework of Ref.~\cite{tandeitnik2026vectorial}, implemented in Ref.~\cite{tandeitnik2026levitationtoolbox}. We summarize what enters it and how its parameters are fixed \emph{in situ}.

\paragraph{Geometric factor.} The trapping field is constructed in the angular spectrum representation~\cite{richards1959electromagnetic,wolf1959electromagnetic,novotny2012principles} for an aplanatic objective of numerical aperture NA, filled by a Gaussian beam of radius $w_0$ and characterized by the filling factor $f_0 = w_0/(f\,\mathrm{NA})$, with $f$ the focal length. The particle radiates as a point dipole, and the detected signal is the overlap $S(z)$ of its far field, collimated by the objective, with the LO mode. Here that mode is the fundamental mode of the collection fibre propagated back to the pupil (fibre core radius \SI{4.1}{\micro\meter}, NA 0.14, coupling lens focal length \SI{3.1}{\milli\meter}). The geometric factor is the phase slope of this overlap,
\begin{equation}\label{eq:kappaDef}
    \kappa = \frac{1}{2k}\,\frac{\partial\arg S(z)}{\partial z}\bigg|_{z=z_{\mathrm{eq}}},
\end{equation}
evaluated at the equilibrium position $z_{\mathrm{eq}}$ of the particle. In the paraxial limit $\kappa = 1$. At finite numerical aperture the average of $\cos\theta$ over the focusing and collection cones makes $\kappa < 1$.

\begin{figure}[tbp]
    \centering
    \fig{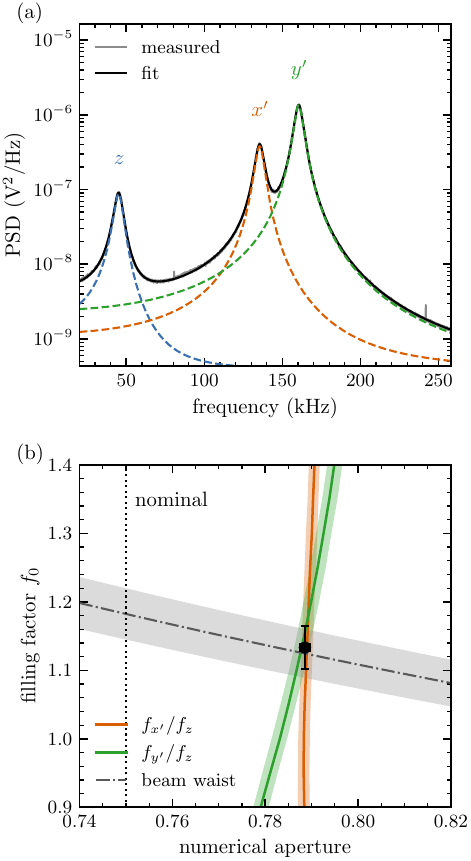}
    \caption{Mechanical frequencies and the effective numerical aperture. (a)~Power spectral density of the forward-detection signal, with a fit of three damped-oscillator lines (dashed) and a white floor. The knife edge sees a projection of both transverse modes, $x'$ and $y'$. (b)~Numerical aperture and filling factor for which the vectorial model reproduces the measured ratios $f_{x'}/f_z$ and $f_{y'}/f_z$ (bands include a \SI{0.3}{\percent} model systematic), and the constraint from the measured input beam (grey band). They cross at $\mathrm{NA} = 0.791$ and $f_0 = 1.13$. The dotted line marks the nominal aperture of the objective.}
    \label{fig:trapFreq}
\end{figure}

\paragraph{Effective numerical aperture.} The ratios of the three mechanical frequencies depend only on the shape of the focus, not on the optical power or the particle, so they fix the parameters of the focal field [Fig.~\ref{fig:trapFreq}]. With the nominal numerical aperture of the objective, 0.75, the model cannot reproduce the measured ratios for any filling factor. A fit of NA and $f_0$, with the measured input beam as a further constraint, gives $\mathrm{NA} = 0.791(3)$ and $f_0 = 1.13(3)$ [Fig.~\ref{fig:trapFreq}(b)]. This lies between the nominal value, which corresponds to the specified clear aperture of the aspheric surface (diameter \SI{0.79}{mm}), and 0.85, set by its full diameter (\SI{0.90}{mm}). The input beam overfills the clear aperture, and the outer zone of the lens contributes to the focus. The two frequency ratios and the beam constraint are consistent within the model, so aberrations of this zone are not resolved at this level. The spectrum is taken just before each measurement, so that $\kappa$ refers to the focus of that run.

\paragraph{Uncertainty.} Of the \SI{0.15}{\percent} uncertainty of $\kappa$ (Sec.~\ref{sec:absolute}), the numerical aperture contributes \SI{0.13}{\percent} and the filling factor \SI{0.08}{\percent}. The nominal numerical aperture would have given a $\kappa$ \SI{1.8}{\percent} higher, so the \emph{in situ} determination of the focal field is necessary at this level of accuracy. Over the excursions used here, up to \SI{450}{nm}, the local slope of $\arg S$ changes by less than \SI{0.2}{\percent}, which biases the thermal variance and the drive amplitude by less than \SI{0.05}{\percent}. \REV{Over $\pm\lambda$ around $z_{\mathrm{eq}}$ the local slope rises to 0.872 and 0.877, the phase departs from a straight line by at most \SI{0.08}{rad} out of \SI{10.8}{rad}, and $|S|$ falls to 0.60 of its value at $z_{\mathrm{eq}}$, so that the phase noise rises by at most a factor of 2.8.}

\section{FPGA demodulation and phase scale}\label{app:fpga}

The demodulation board, a Red Pitaya STEMlab 125-14 (Xilinx Zynq-7010, two 14-bit analog-to-digital and digital-to-analog converters at \SI{125}{\mega\sample\per\second}), digitizes the heterodyne and homodyne signals on the same clock. A numerically controlled oscillator (NCO) provides the references $\cos(\Delta\omega t)$ and $-\sin(\Delta\omega t)$. It is not referenced to the AOM driver, and a frequency-locked loop acting on the demodulated phase removes their mismatch (about \SI{115}{Hz} at \SI{10}{\mega\hertz}) with a bandwidth of about \SI{1}{Hz}. After mixing, a cascaded integrator-comb (CIC) filter~\cite{hogenauer1981} with decimation $D=16$ suppresses the image at $2\Delta\omega$ by about \SI{72}{\decibel}, giving an output rate of \SI{7.8}{\mega\sample\per\second}. A CORDIC core~\cite{volder1959} returns the wrapped phase, with $2^{11}$ counts per radian, and the amplitude $|Q+iI|$. The phase is unwrapped, a first-order filter with a \SI{9.5}{Hz} corner removes the slow drift of $\phi_{\mathrm{LO}}$, and a configurable window, at bit position $l$, selects the bits that are recorded. The records are read digitally by direct memory access (DMA), with the heterodyne phase and the homodyne sample in the same word. The two are therefore simultaneous up to a fixed pipeline delay of about \SI{910}{\nano\second}, which is measured and removed in the analysis.

\begin{figure}[tbp]
    \centering
    \fig{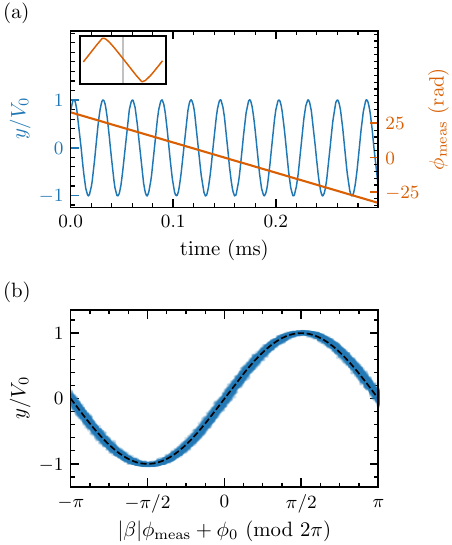}
    \caption{Verification of the digital phase scale with the trap empty and a triangular ramp on a fibre stretcher in the SF path. (a)~Homodyne fringes (left axis) and heterodyne phase (right axis) recorded simultaneously over \SI{0.3}{ms} in the middle of a half-ramp; inset, heterodyne phase over one ramp period, with the window shown shaded. (b)~Homodyne signal of 31 half-ramps (about 4000 fringes) against the heterodyne phase scaled by the fitted $|\beta|$ and folded modulo $2\pi$, each half-ramp aligned by its fitted phase offset, with a sine as the dashed curve. $|\beta| = 0.99937$ at $l = 1$.}
    \label{fig:ramp}
\end{figure}

\paragraph{Verification of the phase scale.}
By construction, one least significant bit of the recorded phase word is
\begin{equation}\label{eq:lsb}
    \delta\phi_{\mathrm{LSB}} = 2^{\,l-11}\ \mathrm{rad},
\end{equation}
with $l$ the position of the window. Since the absolute calibration rests on this scaling, we verify it against an optical ruler. With the trap empty, a fibre stretcher on the SF fibre ramps the phase of a weak back-reflection, the only field present, over about 130 fringes per half-ramp, and both branches record it simultaneously [Fig.~\ref{fig:ramp}(a)]. Consecutive homodyne fringes are separated by exactly $2\pi$ of optical phase. A fit of the homodyne with Eq.~\eqref{eq:transfer} therefore tests the heterodyne scale, with $|\beta| = 1$ if it is exact. Once the known digital filtering is inverted, the fringes of all half-ramps collapse onto a single sinusoid [Fig.~\ref{fig:ramp}(b)]. We obtain $|\beta| = 0.99937$ at two window positions, $l = 1$ and 2. Their ratio, $1.000005(5)$, confirms the factor $2^{-l}$. The remaining $6\times10^{-4}$, common to both, is attributed to the birefringence of the stretched fibre. The digital phase scale is thus verified to better than $10^{-3}$, well below the uncertainty of $\kappa$.

\section{Noise budgets of homodyne and heterodyne readout}\label{app:noise}

\paragraph{Two-quadrature penalty.} A heterodyne measurement records both quadratures of the field, but only one carries position information. The vacuum fluctuations of the other double the imprecision at fixed collected power, so the detection efficiency of a heterodyne is at most $1/2$ before any technical loss.

\paragraph{Dynamic range.} In balanced homodyne detection the strong interference term cancels, and the detector sees only the small signal $V_0\sin\phi$. In heterodyne detection the whole beat carrier at $\Delta\omega$ appears at the output and must fit within the range of the detector and the digitizer. This limits the LO power, and with it the ratio of shot noise to electronic noise, which then no longer grows with the LO power and falls as more scattered light is collected. A high-gain receiver suited to homodyne detection therefore cannot reach the shot-noise limit in heterodyne.

\paragraph{Measured budgets.} Both branches use the same detector model. At the mechanical frequency the optical part of the floor is \SI{47}{\percent} for the homodyne but only \SI{12}{\percent} for the heterodyne, although the heterodyne branch receives \SI{90}{\percent} of the LO. Without electronic noise the homodyne efficiency would rise from 0.08 to 0.17, and the heterodyne efficiency from $1.5\times10^{-3}$ to $6.5\times10^{-3}$.

\section{Parasitic fields}\label{app:parasitic}

\begin{figure}[tbp]
    \centering
    \fig{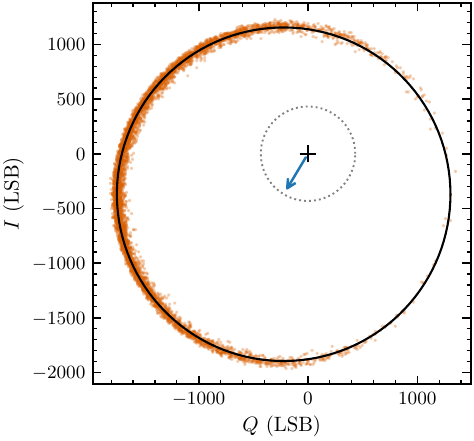}
    \caption{Demodulated I/Q vector of one record with a particle trapped (orange points), with a circle fit (black). The thermal motion moves the vector along an arc of radius $\mathcal{E}_{\mathrm{SF}}$ centred on the parasitic phasor (arrow), not on the origin (cross), here with $r = 3.5$ and $\Delta \approx \SI{-1.2}{rad}$. The dotted circle is the parasitic amplitude measured with the trap empty.}
    \label{fig:iqPlane}
\end{figure}

Collecting the back-scattered light requires an optical circulator that separates it from the incoming trapping beam, in our case a Faraday rotator and a polarizing beam splitter. Any reflection of the trapping beam on the trap side of the circulator returns along the same path and reaches the detector together with the SF, whether the light is collected into a fibre or in free space~\cite{magrini2021real,tebbenjohanns2021quantum}. We consider one parasitic field of amplitude $\mathcal{E}_{\mathrm{PF}}$ and slowly drifting phase $\phi_{\mathrm{PF}}$, and write $r = \mathcal{E}_{\mathrm{SF}}/\mathcal{E}_{\mathrm{PF}}$.

In the homodyne the parasitic field adds a quasi-static offset, which does not distort the response but shifts the lock point as $\phi_{\mathrm{PF}}$ drifts. In the heterodyne it adds a phasor to the demodulated vector, $Q+iI\propto\mathcal{E}_{\mathrm{SF}}e^{i\psi_{\mathrm{SF}}}+\mathcal{E}_{\mathrm{PF}}e^{i\psi_{\mathrm{PF}}}$, with $\psi_{\mathrm{SF}} = \phi_{\mathrm{SF}}-\phi_{\mathrm{LO}}$ and $\psi_{\mathrm{PF}} = \phi_{\mathrm{PF}}-\phi_{\mathrm{LO}}$, and the measured phase becomes
\begin{subequations}\label{eq:phasePF}
    \begin{align}
        \phi_{\mathrm{meas}} &= \psi_{\mathrm{PF}} + \arctan\frac{r\sin\Delta}{1+r\cos\Delta},\\
        \Delta &= \psi_{\mathrm{SF}}-\psi_{\mathrm{PF}}.
    \end{align}
\end{subequations}
The vector traces an arc of radius $\mathcal{E}_{\mathrm{SF}}$ centred on the parasitic phasor instead of the origin [Fig.~\ref{fig:iqPlane}], and the map from $\phi_{\mathrm{SF}}$ to $\phi_{\mathrm{meas}}$ is nonlinear, with a slope $1-\cos\Delta/r$ to first order for $r\gg1$. Averaged over a $\Delta$ that sweeps many radians the mean slope is unity, but for a slowly drifting $\Delta$ a single run can be off by an amount of order $1/r$. Unlike in the homodyne, the parasite is observable: the centre of the arc gives $r$ directly.

We find $r = 3.5$ to 4.7, consistent with the run-to-run spread of $\beta$ (Sec.~\ref{sec:linearity}). The scale factor $\beta$ corrects the mean slope of the distorted map but not its curvature, and masses computed from the $\beta$-corrected phase scattered by \SIrange{3}{12}{\percent} between runs. The mass of Sec.~\ref{sec:mass} avoids this: a circle fit to the arc locates its centre, and the phase about the centre is $\phi_{\mathrm{SF}}$ itself, for any $\Delta$. Parasitic fields can also be suppressed optically, by tilting interfaces and by confocal filtering of the collected light~\cite{magrini2021real}.

\section{Nonlinear bias of thermometric homodyne calibration}\label{app:bussgang}

Let the motional phase $\phi(t)$ be a stationary Gaussian process with zero mean, variance $\sigma_\phi^2$ and autocorrelation $R(\tau)$, and let the homodyne output be $y = V_0\sin(\phi+\phi_0)$. For a Gaussian input and a memoryless nonlinearity, Bussgang's theorem~\cite{bussgang1952crosscorrelation} gives $\langle y(t)\phi(t+\tau)\rangle = G\,R(\tau)$, with
\begin{equation}\label{eq:bussgangGain}
    G = \langle\partial y/\partial\phi\rangle = V_0\cos\phi_0\,e^{-\sigma_\phi^2/2}.
\end{equation}
The output thus splits into a component $G\phi(t)$ and a distortion uncorrelated with $\phi$. Its autocorrelation is
\begin{equation}\label{eq:homodyneAutocorr}
    \begin{split}
    \langle y(t)y(t+\tau)\rangle = V_0^2e^{-\sigma_\phi^2}\big[&\cos^2\phi_0\sinh R(\tau)\\
    &+ \sin^2\phi_0\big(\cosh R(\tau)-1\big)\big],
    \end{split}
\end{equation}
whose leading term is the linear part $G^2R(\tau)$, while higher powers of $R$ produce the harmonics. A Lorentzian fit to the homodyne PSD returns the linear part only, so the thermometric coefficient is $2k_{\mathrm{eff}}G$, which gives Eq.~\eqref{eq:bussgang}.

With a coherent drive the homodyne records $y(t) = V_0\sin(\phi_A\sin\omega_{\mathrm{d}}t + \phi_{\mathrm{th}}(t) + \phi_0)$, with a Gaussian thermal phase $\phi_{\mathrm{th}}$ that is independent of the drive, since the oscillator is linear. Lock-in detection averages over $\phi_{\mathrm{th}}$, and since $\langle\cos\phi_{\mathrm{th}}\rangle = e^{-\sigma_\phi^2/2}$ and $\langle\sin\phi_{\mathrm{th}}\rangle = 0$, the average output is Eq.~\eqref{eq:coherent}. With $\theta = \omega_{\mathrm{d}}t$, the Jacobi--Anger expansion gives
\begin{equation}\label{eq:jacobiAngerOffset}
    \begin{split}
    &\sin(\phi_A\sin\theta + \phi_0)\\
    &\quad= 2\cos\phi_0\sum_{k=0}^{\infty}J_{2k+1}(\phi_A)\sin\big((2k+1)\theta\big)\\
    &\quad\phantom{=}+ \sin\phi_0\Big[J_0(\phi_A) + 2\sum_{k=1}^{\infty}J_{2k}(\phi_A)\cos(2k\theta)\Big],
    \end{split}
\end{equation}
so the odd harmonics carry $\cos\phi_0$ and give Eq.~\eqref{eq:Yn}. The constant term, proportional to $J_0(\phi_A)\sin\phi_0$, is the signal on which the phase lock acts, and it changes sign at the first zero of $J_0$ (Sec.~\ref{sec:driven}).

\section{Absorption heating of the particle}\label{app:heating}

\paragraph{Two-bath temperature.} In the free-molecular regime the centre of mass couples to the gas molecules impinging at $T$ and to those emitted from the hot surface at $T_{\mathrm{em}}$~\cite{millen2014nanoscale},
\begin{subequations}\label{eq:twoBath}
    \begin{align}
        T_{\mathrm{cm}} &= \frac{T + \zeta\,T_{\mathrm{em}}}{1+\zeta},\\
        \zeta &= \frac{\pi}{8}\sqrt{\frac{T_{\mathrm{em}}}{T}}.
    \end{align}
\end{subequations}
Liquid water absorbs \SI{9.7}{cm^{-1}} at \SI{1550}{nm} against \SI{0.12}{cm^{-1}} at \SI{1064}{nm}~\cite{hale1973optical}, which is why the heating, small at \SI{1064}{nm}~\cite{hebestreit2018internal}, is visible at \SI{1550}{nm} at the pressures where calibrations are taken.

\paragraph{Form of the heating law.} Equation~\eqref{eq:heating} is the first order of a model in which a constant absorbed power, balanced by free-molecular conduction, enters $T_{\mathrm{cm}}$ through Eq.~\eqref{eq:twoBath}. This model fits every series as well as the linear law, and moves $m$ by less than \SI{0.05}{\percent} after conditioning, but by \SIrange{0.5}{1.3}{\percent} as trapped.

\paragraph{Mass loss and re-adsorption.} Conditioning lowers $m$ by \SIrange{0.6}{5}{\percent} (Table~\ref{tab:mass}), and the particles with the largest heating coefficient as trapped lose the most. A loss of \SI{1}{\percent} corresponds to about \SI{0.5}{nm} of water on the surface of a \SI{150}{nm} sphere, one to two monolayers. During the re-adsorption of Fig.~\ref{fig:massCampaign}(c), followed for \SI{64}{h} with the pump off while leaks raised the pressure from \SIrange{6}{21}{mbar}, $m$ grows by \SI{2}{\percent} with a time constant of \SI{13(2)}{h} and $c$ by a factor of 2.5 with \SI{16(3)}{h}. Both level off below their values as trapped. This agrees with the partly reversible dehydration of silica nanoparticles observed by Ricci \emph{et al.}~\cite{ricci2022chemical}, who also found an irreversible dehydroxylation, with a mass loss of \SI{9}{\percent}, above a bulk temperature of \SIrange{400}{860}{K}. Our conditioning does not reach that state: the mass changes by less, and the particles take up water again.

\paragraph{Single spheres.} The gas damping of the three modes is isotropic for a sphere, whereas a dimer has damping rates differing by about \SI{27}{\percent} along and across its axis~\cite{ahn2018optically}. For the five particles the ratios of the transverse to the axial damping lie between 0.95 and 1.05.

\bibliography{main}

\begin{thebibliography}{43}%
\makeatletter
\providecommand \@ifxundefined [1]{%
 \@ifx{#1\undefined}
}%
\providecommand \@ifnum [1]{%
 \ifnum #1\expandafter \@firstoftwo
 \else \expandafter \@secondoftwo
 \fi
}%
\providecommand \@ifx [1]{%
 \ifx #1\expandafter \@firstoftwo
 \else \expandafter \@secondoftwo
 \fi
}%
\providecommand \natexlab [1]{#1}%
\providecommand \enquote  [1]{``#1''}%
\providecommand \bibnamefont  [1]{#1}%
\providecommand \bibfnamefont [1]{#1}%
\providecommand \citenamefont [1]{#1}%
\providecommand \href@noop [0]{\@secondoftwo}%
\providecommand \href [0]{\begingroup \@sanitize@url \@href}%
\providecommand \@href[1]{\@@startlink{#1}\@@href}%
\providecommand \@@href[1]{\endgroup#1\@@endlink}%
\providecommand \@sanitize@url [0]{\catcode `\\12\catcode `\$12\catcode
  `\&12\catcode `\#12\catcode `\^12\catcode `\_12\catcode `\%12\relax}%
\providecommand \@@startlink[1]{}%
\providecommand \@@endlink[0]{}%
\providecommand \url  [0]{\begingroup\@sanitize@url \@url }%
\providecommand \@url [1]{\endgroup\@href {#1}{\urlprefix }}%
\providecommand \urlprefix  [0]{URL }%
\providecommand \Eprint [0]{\href }%
\providecommand \doibase [0]{https://doi.org/}%
\providecommand \selectlanguage [0]{\@gobble}%
\providecommand \bibinfo  [0]{\@secondoftwo}%
\providecommand \bibfield  [0]{\@secondoftwo}%
\providecommand \translation [1]{[#1]}%
\providecommand \BibitemOpen [0]{}%
\providecommand \bibitemStop [0]{}%
\providecommand \bibitemNoStop [0]{.\EOS\space}%
\providecommand \EOS [0]{\spacefactor3000\relax}%
\providecommand \BibitemShut  [1]{\csname bibitem#1\endcsname}%
\let\auto@bib@innerbib\@empty
%</preamble>
\bibitem [{\citenamefont {Braginsky}\ and\ \citenamefont
  {Khalili}(1995)}]{braginsky1995quantum}%
  \BibitemOpen
  \bibfield  {author} {\bibinfo {author} {\bibfnamefont {V.~B.}\ \bibnamefont
  {Braginsky}}\ and\ \bibinfo {author} {\bibfnamefont {F.~Y.}\ \bibnamefont
  {Khalili}},\ }\href@noop {} {\emph {\bibinfo {title} {Quantum measurement}}}\
  (\bibinfo  {publisher} {Cambridge University Press},\ \bibinfo {year}
  {1995})\BibitemShut {NoStop}%
\bibitem [{\citenamefont {{LIGO Scientific Collaboration}}\ \emph
  {et~al.}(2015)\citenamefont {{LIGO Scientific Collaboration}}, \citenamefont
  {Aasi}, \citenamefont {Abbott}, \citenamefont {Abbott}, \citenamefont
  {Abbott}, \citenamefont {Abernathy} \emph {et~al.}}]{aasi2015advancedligo}%
  \BibitemOpen
  \bibfield  {author} {\bibinfo {author} {\bibnamefont {{LIGO Scientific
  Collaboration}}}, \bibinfo {author} {\bibfnamefont {J.}~\bibnamefont {Aasi}},
  \bibinfo {author} {\bibfnamefont {B.~P.}\ \bibnamefont {Abbott}}, \bibinfo
  {author} {\bibfnamefont {R.}~\bibnamefont {Abbott}}, \bibinfo {author}
  {\bibfnamefont {T.}~\bibnamefont {Abbott}}, \bibinfo {author} {\bibfnamefont
  {M.~R.}\ \bibnamefont {Abernathy}}, \emph {et~al.},\ }\bibfield  {title}
  {\bibinfo {title} {Advanced {LIGO}},\ }\href
  {https://doi.org/10.1088/0264-9381/32/7/074001} {\bibfield  {journal}
  {\bibinfo  {journal} {Classical and Quantum Gravity}\ }\textbf {\bibinfo
  {volume} {32}},\ \bibinfo {pages} {074001} (\bibinfo {year}
  {2015})}\BibitemShut {NoStop}%
\bibitem [{\citenamefont {Abbott}\ \emph {et~al.}(2016)\citenamefont {Abbott},
  \citenamefont {Abbott}, \citenamefont {Abbott}, \citenamefont {Abernathy},
  \citenamefont {Acernese}, \citenamefont {Ackley}, \citenamefont {Adams},
  \citenamefont {Adams}, \citenamefont {Addesso}, \citenamefont {Adhikari}
  \emph {et~al.}}]{abbott2016observation}%
  \BibitemOpen
  \bibfield  {author} {\bibinfo {author} {\bibfnamefont {B.~P.}\ \bibnamefont
  {Abbott}}, \bibinfo {author} {\bibfnamefont {R.}~\bibnamefont {Abbott}},
  \bibinfo {author} {\bibfnamefont {T.~D.}\ \bibnamefont {Abbott}}, \bibinfo
  {author} {\bibfnamefont {M.~R.}\ \bibnamefont {Abernathy}}, \bibinfo {author}
  {\bibfnamefont {F.}~\bibnamefont {Acernese}}, \bibinfo {author}
  {\bibfnamefont {K.}~\bibnamefont {Ackley}}, \bibinfo {author} {\bibfnamefont
  {C.}~\bibnamefont {Adams}}, \bibinfo {author} {\bibfnamefont
  {T.}~\bibnamefont {Adams}}, \bibinfo {author} {\bibfnamefont
  {P.}~\bibnamefont {Addesso}}, \bibinfo {author} {\bibfnamefont {R.~X.}\
  \bibnamefont {Adhikari}}, \emph {et~al.},\ }\bibfield  {title} {\bibinfo
  {title} {Observation of gravitational waves from a binary black hole
  merger},\ }\href@noop {} {\bibfield  {journal} {\bibinfo  {journal} {Physical
  review letters}\ }\textbf {\bibinfo {volume} {116}},\ \bibinfo {pages}
  {061102} (\bibinfo {year} {2016})}\BibitemShut {NoStop}%
\bibitem [{\citenamefont {Aspelmeyer}\ \emph {et~al.}(2014)\citenamefont
  {Aspelmeyer}, \citenamefont {Kippenberg},\ and\ \citenamefont
  {Marquardt}}]{aspelmeyer2014cavity}%
  \BibitemOpen
  \bibfield  {author} {\bibinfo {author} {\bibfnamefont {M.}~\bibnamefont
  {Aspelmeyer}}, \bibinfo {author} {\bibfnamefont {T.~J.}\ \bibnamefont
  {Kippenberg}},\ and\ \bibinfo {author} {\bibfnamefont {F.}~\bibnamefont
  {Marquardt}},\ }\bibfield  {title} {\bibinfo {title} {Cavity optomechanics},\
  }\href {https://doi.org/10.1103/RevModPhys.86.1391} {\bibfield  {journal}
  {\bibinfo  {journal} {Reviews of Modern Physics}\ }\textbf {\bibinfo {volume}
  {86}},\ \bibinfo {pages} {1391} (\bibinfo {year} {2014})}\BibitemShut
  {NoStop}%
\bibitem [{\citenamefont {Magrini}\ \emph {et~al.}(2021)\citenamefont
  {Magrini}, \citenamefont {Rosenzweig}, \citenamefont {Bach}, \citenamefont
  {Deutschmann-Olek}, \citenamefont {Hofer}, \citenamefont {Hong},
  \citenamefont {Kiesel}, \citenamefont {Kugi},\ and\ \citenamefont
  {Aspelmeyer}}]{magrini2021real}%
  \BibitemOpen
  \bibfield  {author} {\bibinfo {author} {\bibfnamefont {L.}~\bibnamefont
  {Magrini}}, \bibinfo {author} {\bibfnamefont {P.}~\bibnamefont {Rosenzweig}},
  \bibinfo {author} {\bibfnamefont {C.}~\bibnamefont {Bach}}, \bibinfo {author}
  {\bibfnamefont {A.}~\bibnamefont {Deutschmann-Olek}}, \bibinfo {author}
  {\bibfnamefont {S.~G.}\ \bibnamefont {Hofer}}, \bibinfo {author}
  {\bibfnamefont {S.}~\bibnamefont {Hong}}, \bibinfo {author} {\bibfnamefont
  {N.}~\bibnamefont {Kiesel}}, \bibinfo {author} {\bibfnamefont
  {A.}~\bibnamefont {Kugi}},\ and\ \bibinfo {author} {\bibfnamefont
  {M.}~\bibnamefont {Aspelmeyer}},\ }\bibfield  {title} {\bibinfo {title}
  {Real-time optimal quantum control of mechanical motion at room
  temperature},\ }\href@noop {} {\bibfield  {journal} {\bibinfo  {journal}
  {Nature}\ }\textbf {\bibinfo {volume} {595}},\ \bibinfo {pages} {373}
  (\bibinfo {year} {2021})}\BibitemShut {NoStop}%
\bibitem [{\citenamefont {Tebbenjohanns}\ \emph {et~al.}(2021)\citenamefont
  {Tebbenjohanns}, \citenamefont {Mattana}, \citenamefont {Rossi},
  \citenamefont {Frimmer},\ and\ \citenamefont
  {Novotny}}]{tebbenjohanns2021quantum}%
  \BibitemOpen
  \bibfield  {author} {\bibinfo {author} {\bibfnamefont {F.}~\bibnamefont
  {Tebbenjohanns}}, \bibinfo {author} {\bibfnamefont {M.~L.}\ \bibnamefont
  {Mattana}}, \bibinfo {author} {\bibfnamefont {M.}~\bibnamefont {Rossi}},
  \bibinfo {author} {\bibfnamefont {M.}~\bibnamefont {Frimmer}},\ and\ \bibinfo
  {author} {\bibfnamefont {L.}~\bibnamefont {Novotny}},\ }\bibfield  {title}
  {\bibinfo {title} {Quantum control of a nanoparticle optically levitated in
  cryogenic free space},\ }\href@noop {} {\bibfield  {journal} {\bibinfo
  {journal} {Nature}\ }\textbf {\bibinfo {volume} {595}},\ \bibinfo {pages}
  {378} (\bibinfo {year} {2021})}\BibitemShut {NoStop}%
\bibitem [{\citenamefont {Iwasaki}\ \emph {et~al.}(2019)\citenamefont
  {Iwasaki}, \citenamefont {Yotsuya}, \citenamefont {Naruki}, \citenamefont
  {Matsuda}, \citenamefont {Yoneda},\ and\ \citenamefont
  {Aikawa}}]{iwasaki2019electric}%
  \BibitemOpen
  \bibfield  {author} {\bibinfo {author} {\bibfnamefont {M.}~\bibnamefont
  {Iwasaki}}, \bibinfo {author} {\bibfnamefont {T.}~\bibnamefont {Yotsuya}},
  \bibinfo {author} {\bibfnamefont {T.}~\bibnamefont {Naruki}}, \bibinfo
  {author} {\bibfnamefont {Y.}~\bibnamefont {Matsuda}}, \bibinfo {author}
  {\bibfnamefont {M.}~\bibnamefont {Yoneda}},\ and\ \bibinfo {author}
  {\bibfnamefont {K.}~\bibnamefont {Aikawa}},\ }\bibfield  {title} {\bibinfo
  {title} {Electric feedback cooling of single charged nanoparticles in an
  optical trap},\ }\href@noop {} {\bibfield  {journal} {\bibinfo  {journal}
  {Physical Review A}\ }\textbf {\bibinfo {volume} {99}},\ \bibinfo {pages}
  {051401} (\bibinfo {year} {2019})}\BibitemShut {NoStop}%
\bibitem [{\citenamefont {Kremer}\ \emph {et~al.}(2024)\citenamefont {Kremer},
  \citenamefont {Califrer}, \citenamefont {Tandeitnik}, \citenamefont {von~der
  Weid}, \citenamefont {Tempor{\~a}o},\ and\ \citenamefont
  {Guerreiro}}]{kremer2024all}%
  \BibitemOpen
  \bibfield  {author} {\bibinfo {author} {\bibfnamefont {O.}~\bibnamefont
  {Kremer}}, \bibinfo {author} {\bibfnamefont {I.}~\bibnamefont {Califrer}},
  \bibinfo {author} {\bibfnamefont {D.}~\bibnamefont {Tandeitnik}}, \bibinfo
  {author} {\bibfnamefont {J.~P.}\ \bibnamefont {von~der Weid}}, \bibinfo
  {author} {\bibfnamefont {G.}~\bibnamefont {Tempor{\~a}o}},\ and\ \bibinfo
  {author} {\bibfnamefont {T.}~\bibnamefont {Guerreiro}},\ }\bibfield  {title}
  {\bibinfo {title} {All-electrical cooling of an optically levitated
  nanoparticle},\ }\href@noop {} {\bibfield  {journal} {\bibinfo  {journal}
  {Physical Review Applied}\ }\textbf {\bibinfo {volume} {22}},\ \bibinfo
  {pages} {024010} (\bibinfo {year} {2024})}\BibitemShut {NoStop}%
\bibitem [{\citenamefont {Thompson}\ \emph {et~al.}(2008)\citenamefont
  {Thompson}, \citenamefont {Zwickl}, \citenamefont {Jayich}, \citenamefont
  {Marquardt}, \citenamefont {Girvin},\ and\ \citenamefont
  {Harris}}]{thompson2008strong}%
  \BibitemOpen
  \bibfield  {author} {\bibinfo {author} {\bibfnamefont {J.}~\bibnamefont
  {Thompson}}, \bibinfo {author} {\bibfnamefont {B.}~\bibnamefont {Zwickl}},
  \bibinfo {author} {\bibfnamefont {A.}~\bibnamefont {Jayich}}, \bibinfo
  {author} {\bibfnamefont {F.}~\bibnamefont {Marquardt}}, \bibinfo {author}
  {\bibfnamefont {S.}~\bibnamefont {Girvin}},\ and\ \bibinfo {author}
  {\bibfnamefont {J.}~\bibnamefont {Harris}},\ }\bibfield  {title} {\bibinfo
  {title} {Strong dispersive coupling of a high-finesse cavity to a
  micromechanical membrane},\ }\href@noop {} {\bibfield  {journal} {\bibinfo
  {journal} {Nature}\ }\textbf {\bibinfo {volume} {452}},\ \bibinfo {pages}
  {72} (\bibinfo {year} {2008})}\BibitemShut {NoStop}%
\bibitem [{\citenamefont {Chan}\ \emph {et~al.}(2011)\citenamefont {Chan},
  \citenamefont {Alegre}, \citenamefont {Safavi-Naeini}, \citenamefont {Hill},
  \citenamefont {Krause}, \citenamefont {Gr{\"o}blacher}, \citenamefont
  {Aspelmeyer},\ and\ \citenamefont {Painter}}]{chan2011laser}%
  \BibitemOpen
  \bibfield  {author} {\bibinfo {author} {\bibfnamefont {J.}~\bibnamefont
  {Chan}}, \bibinfo {author} {\bibfnamefont {T.~M.}\ \bibnamefont {Alegre}},
  \bibinfo {author} {\bibfnamefont {A.~H.}\ \bibnamefont {Safavi-Naeini}},
  \bibinfo {author} {\bibfnamefont {J.~T.}\ \bibnamefont {Hill}}, \bibinfo
  {author} {\bibfnamefont {A.}~\bibnamefont {Krause}}, \bibinfo {author}
  {\bibfnamefont {S.}~\bibnamefont {Gr{\"o}blacher}}, \bibinfo {author}
  {\bibfnamefont {M.}~\bibnamefont {Aspelmeyer}},\ and\ \bibinfo {author}
  {\bibfnamefont {O.}~\bibnamefont {Painter}},\ }\bibfield  {title} {\bibinfo
  {title} {Laser cooling of a nanomechanical oscillator into its quantum ground
  state},\ }\href@noop {} {\bibfield  {journal} {\bibinfo  {journal} {Nature}\
  }\textbf {\bibinfo {volume} {478}},\ \bibinfo {pages} {89} (\bibinfo {year}
  {2011})}\BibitemShut {NoStop}%
\bibitem [{\citenamefont {Gavartin}\ \emph {et~al.}(2012)\citenamefont
  {Gavartin}, \citenamefont {Verlot},\ and\ \citenamefont
  {Kippenberg}}]{gavartin2012hybrid}%
  \BibitemOpen
  \bibfield  {author} {\bibinfo {author} {\bibfnamefont {E.}~\bibnamefont
  {Gavartin}}, \bibinfo {author} {\bibfnamefont {P.}~\bibnamefont {Verlot}},\
  and\ \bibinfo {author} {\bibfnamefont {T.~J.}\ \bibnamefont {Kippenberg}},\
  }\bibfield  {title} {\bibinfo {title} {A hybrid on-chip optomechanical
  transducer for ultrasensitive force measurements},\ }\href@noop {} {\bibfield
   {journal} {\bibinfo  {journal} {Nature nanotechnology}\ }\textbf {\bibinfo
  {volume} {7}},\ \bibinfo {pages} {509} (\bibinfo {year} {2012})}\BibitemShut
  {NoStop}%
\bibitem [{\citenamefont {Purdy}\ \emph {et~al.}(2013)\citenamefont {Purdy},
  \citenamefont {Peterson},\ and\ \citenamefont
  {Regal}}]{purdy2013observation}%
  \BibitemOpen
  \bibfield  {author} {\bibinfo {author} {\bibfnamefont {T.~P.}\ \bibnamefont
  {Purdy}}, \bibinfo {author} {\bibfnamefont {R.~W.}\ \bibnamefont
  {Peterson}},\ and\ \bibinfo {author} {\bibfnamefont {C.}~\bibnamefont
  {Regal}},\ }\bibfield  {title} {\bibinfo {title} {Observation of radiation
  pressure shot noise on a macroscopic object},\ }\href@noop {} {\bibfield
  {journal} {\bibinfo  {journal} {Science}\ }\textbf {\bibinfo {volume}
  {339}},\ \bibinfo {pages} {801} (\bibinfo {year} {2013})}\BibitemShut
  {NoStop}%
\bibitem [{\citenamefont {Clerk}\ \emph {et~al.}(2010)\citenamefont {Clerk},
  \citenamefont {Devoret}, \citenamefont {Girvin}, \citenamefont {Marquardt},\
  and\ \citenamefont {Schoelkopf}}]{clerk2010introduction}%
  \BibitemOpen
  \bibfield  {author} {\bibinfo {author} {\bibfnamefont {A.~A.}\ \bibnamefont
  {Clerk}}, \bibinfo {author} {\bibfnamefont {M.~H.}\ \bibnamefont {Devoret}},
  \bibinfo {author} {\bibfnamefont {S.~M.}\ \bibnamefont {Girvin}}, \bibinfo
  {author} {\bibfnamefont {F.}~\bibnamefont {Marquardt}},\ and\ \bibinfo
  {author} {\bibfnamefont {R.~J.}\ \bibnamefont {Schoelkopf}},\ }\bibfield
  {title} {\bibinfo {title} {Introduction to quantum noise, measurement, and
  amplification},\ }\href@noop {} {\bibfield  {journal} {\bibinfo  {journal}
  {Reviews of Modern Physics}\ }\textbf {\bibinfo {volume} {82}},\ \bibinfo
  {pages} {1155} (\bibinfo {year} {2010})}\BibitemShut {NoStop}%
\bibitem [{\citenamefont {Tebbenjohanns}\ \emph {et~al.}(2019)\citenamefont
  {Tebbenjohanns}, \citenamefont {Frimmer},\ and\ \citenamefont
  {Novotny}}]{tebbenjohanns2019optimal}%
  \BibitemOpen
  \bibfield  {author} {\bibinfo {author} {\bibfnamefont {F.}~\bibnamefont
  {Tebbenjohanns}}, \bibinfo {author} {\bibfnamefont {M.}~\bibnamefont
  {Frimmer}},\ and\ \bibinfo {author} {\bibfnamefont {L.}~\bibnamefont
  {Novotny}},\ }\bibfield  {title} {\bibinfo {title} {Optimal position
  detection of a dipolar scatterer in a focused field},\ }\href@noop {}
  {\bibfield  {journal} {\bibinfo  {journal} {Physical Review A}\ }\textbf
  {\bibinfo {volume} {100}},\ \bibinfo {pages} {043821} (\bibinfo {year}
  {2019})}\BibitemShut {NoStop}%
\bibitem [{\citenamefont {Militaru}\ \emph {et~al.}(2022)\citenamefont
  {Militaru}, \citenamefont {Rossi}, \citenamefont {Tebbenjohanns},
  \citenamefont {Romero-Isart}, \citenamefont {Frimmer},\ and\ \citenamefont
  {Novotny}}]{militaru2022ponderomotive}%
  \BibitemOpen
  \bibfield  {author} {\bibinfo {author} {\bibfnamefont {A.}~\bibnamefont
  {Militaru}}, \bibinfo {author} {\bibfnamefont {M.}~\bibnamefont {Rossi}},
  \bibinfo {author} {\bibfnamefont {F.}~\bibnamefont {Tebbenjohanns}}, \bibinfo
  {author} {\bibfnamefont {O.}~\bibnamefont {Romero-Isart}}, \bibinfo {author}
  {\bibfnamefont {M.}~\bibnamefont {Frimmer}},\ and\ \bibinfo {author}
  {\bibfnamefont {L.}~\bibnamefont {Novotny}},\ }\bibfield  {title} {\bibinfo
  {title} {Ponderomotive squeezing of light by a levitated nanoparticle in free
  space},\ }\href {https://doi.org/10.1103/PhysRevLett.129.053602} {\bibfield
  {journal} {\bibinfo  {journal} {Physical Review Letters}\ }\textbf {\bibinfo
  {volume} {129}},\ \bibinfo {pages} {053602} (\bibinfo {year}
  {2022})}\BibitemShut {NoStop}%
\bibitem [{\citenamefont {Maurer}\ \emph {et~al.}(2023)\citenamefont {Maurer},
  \citenamefont {Gonzalez-Ballestero},\ and\ \citenamefont
  {Romero-Isart}}]{maurer2023quantum}%
  \BibitemOpen
  \bibfield  {author} {\bibinfo {author} {\bibfnamefont {P.}~\bibnamefont
  {Maurer}}, \bibinfo {author} {\bibfnamefont {C.}~\bibnamefont
  {Gonzalez-Ballestero}},\ and\ \bibinfo {author} {\bibfnamefont
  {O.}~\bibnamefont {Romero-Isart}},\ }\bibfield  {title} {\bibinfo {title}
  {Quantum theory of light interaction with a lorenz-mie particle: Optical
  detection and three-dimensional ground-state cooling},\ }\href@noop {}
  {\bibfield  {journal} {\bibinfo  {journal} {Physical Review A}\ }\textbf
  {\bibinfo {volume} {108}},\ \bibinfo {pages} {033714} (\bibinfo {year}
  {2023})}\BibitemShut {NoStop}%
\bibitem [{\citenamefont {Hebestreit}\ \emph
  {et~al.}(2018{\natexlab{a}})\citenamefont {Hebestreit}, \citenamefont
  {Frimmer}, \citenamefont {Reimann}, \citenamefont {Dellago}, \citenamefont
  {Ricci},\ and\ \citenamefont {Novotny}}]{hebestreit2018calibration}%
  \BibitemOpen
  \bibfield  {author} {\bibinfo {author} {\bibfnamefont {E.}~\bibnamefont
  {Hebestreit}}, \bibinfo {author} {\bibfnamefont {M.}~\bibnamefont {Frimmer}},
  \bibinfo {author} {\bibfnamefont {R.}~\bibnamefont {Reimann}}, \bibinfo
  {author} {\bibfnamefont {C.}~\bibnamefont {Dellago}}, \bibinfo {author}
  {\bibfnamefont {F.}~\bibnamefont {Ricci}},\ and\ \bibinfo {author}
  {\bibfnamefont {L.}~\bibnamefont {Novotny}},\ }\bibfield  {title} {\bibinfo
  {title} {Calibration and energy measurement of optically levitated
  nanoparticle sensors},\ }\href {https://doi.org/10.1063/1.5017119} {\bibfield
   {journal} {\bibinfo  {journal} {Review of Scientific Instruments}\ }\textbf
  {\bibinfo {volume} {89}},\ \bibinfo {pages} {033111} (\bibinfo {year}
  {2018}{\natexlab{a}})}\BibitemShut {NoStop}%
\bibitem [{\citenamefont {Ricci}\ \emph {et~al.}(2019)\citenamefont {Ricci},
  \citenamefont {Cuairan}, \citenamefont {Conangla}, \citenamefont {Schell},\
  and\ \citenamefont {Quidant}}]{ricci2019accurate}%
  \BibitemOpen
  \bibfield  {author} {\bibinfo {author} {\bibfnamefont {F.}~\bibnamefont
  {Ricci}}, \bibinfo {author} {\bibfnamefont {M.~T.}\ \bibnamefont {Cuairan}},
  \bibinfo {author} {\bibfnamefont {G.~P.}\ \bibnamefont {Conangla}}, \bibinfo
  {author} {\bibfnamefont {A.~W.}\ \bibnamefont {Schell}},\ and\ \bibinfo
  {author} {\bibfnamefont {R.}~\bibnamefont {Quidant}},\ }\bibfield  {title}
  {\bibinfo {title} {Accurate mass measurement of a levitated nanomechanical
  resonator for precision force-sensing},\ }\href@noop {} {\bibfield  {journal}
  {\bibinfo  {journal} {Nano letters}\ }\textbf {\bibinfo {volume} {19}},\
  \bibinfo {pages} {6711} (\bibinfo {year} {2019})}\BibitemShut {NoStop}%
\bibitem [{\citenamefont {Tandeitnik}\ \emph {et~al.}(2026)\citenamefont
  {Tandeitnik}, \citenamefont {Bianchi}, \citenamefont {Gutierrez-Bernal},
  \citenamefont {Zieli{\'n}ska}, \citenamefont {Neto},\ and\ \citenamefont
  {Guerreiro}}]{tandeitnik2026vectorial}%
  \BibitemOpen
  \bibfield  {author} {\bibinfo {author} {\bibfnamefont {D.}~\bibnamefont
  {Tandeitnik}}, \bibinfo {author} {\bibfnamefont {L.}~\bibnamefont {Bianchi}},
  \bibinfo {author} {\bibfnamefont {S.}~\bibnamefont {Gutierrez-Bernal}},
  \bibinfo {author} {\bibfnamefont {J.~A.}\ \bibnamefont {Zieli{\'n}ska}},
  \bibinfo {author} {\bibfnamefont {P.~A.~M.}\ \bibnamefont {Neto}},\ and\
  \bibinfo {author} {\bibfnamefont {T.}~\bibnamefont {Guerreiro}},\ }\bibfield
  {title} {\bibinfo {title} {Position measurement of a levitated particle with
  vectorial light},\ }\href@noop {} {\bibfield  {journal} {\bibinfo  {journal}
  {arXiv preprint arXiv:2607.23833}\ } (\bibinfo {year} {2026})}\BibitemShut
  {NoStop}%
\bibitem [{\citenamefont {Rider}\ \emph {et~al.}(2018)\citenamefont {Rider},
  \citenamefont {Blakemore}, \citenamefont {Gratta},\ and\ \citenamefont
  {Moore}}]{rider2018}%
  \BibitemOpen
  \bibfield  {author} {\bibinfo {author} {\bibfnamefont {A.~D.}\ \bibnamefont
  {Rider}}, \bibinfo {author} {\bibfnamefont {C.~P.}\ \bibnamefont
  {Blakemore}}, \bibinfo {author} {\bibfnamefont {G.}~\bibnamefont {Gratta}},\
  and\ \bibinfo {author} {\bibfnamefont {D.~C.}\ \bibnamefont {Moore}},\
  }\bibfield  {title} {\bibinfo {title} {Single-beam dielectric-microsphere
  trapping with optical heterodyne detection},\ }\href
  {https://doi.org/10.1103/PhysRevA.97.013842} {\bibfield  {journal} {\bibinfo
  {journal} {Physical Review A}\ }\textbf {\bibinfo {volume} {97}},\ \bibinfo
  {pages} {013842} (\bibinfo {year} {2018})}\BibitemShut {NoStop}%
\bibitem [{\citenamefont {Magrini}\ \emph {et~al.}(2022)\citenamefont
  {Magrini}, \citenamefont {Camarena-Ch{\'a}vez}, \citenamefont {Bach},
  \citenamefont {Johnson},\ and\ \citenamefont
  {Aspelmeyer}}]{magrini2022squeezed}%
  \BibitemOpen
  \bibfield  {author} {\bibinfo {author} {\bibfnamefont {L.}~\bibnamefont
  {Magrini}}, \bibinfo {author} {\bibfnamefont {V.~A.}\ \bibnamefont
  {Camarena-Ch{\'a}vez}}, \bibinfo {author} {\bibfnamefont {C.}~\bibnamefont
  {Bach}}, \bibinfo {author} {\bibfnamefont {A.}~\bibnamefont {Johnson}},\ and\
  \bibinfo {author} {\bibfnamefont {M.}~\bibnamefont {Aspelmeyer}},\ }\bibfield
   {title} {\bibinfo {title} {Squeezed light from a levitated nanoparticle at
  room temperature},\ }\href {https://doi.org/10.1103/PhysRevLett.129.053601}
  {\bibfield  {journal} {\bibinfo  {journal} {Physical Review Letters}\
  }\textbf {\bibinfo {volume} {129}},\ \bibinfo {pages} {053601} (\bibinfo
  {year} {2022})}\BibitemShut {NoStop}%
\bibitem [{\citenamefont {Tseng}\ \emph {et~al.}(2025)\citenamefont {Tseng},
  \citenamefont {Penny}, \citenamefont {Siegel}, \citenamefont {Wang},\ and\
  \citenamefont {Moore}}]{tseng2025search}%
  \BibitemOpen
  \bibfield  {author} {\bibinfo {author} {\bibfnamefont {Y.-H.}\ \bibnamefont
  {Tseng}}, \bibinfo {author} {\bibfnamefont {T.}~\bibnamefont {Penny}},
  \bibinfo {author} {\bibfnamefont {B.}~\bibnamefont {Siegel}}, \bibinfo
  {author} {\bibfnamefont {J.}~\bibnamefont {Wang}},\ and\ \bibinfo {author}
  {\bibfnamefont {D.~C.}\ \bibnamefont {Moore}},\ }\bibfield  {title} {\bibinfo
  {title} {Search for dark matter scattering from optically levitated
  nanoparticles},\ }\href {https://doi.org/10.1103/j76m-gcp1} {\bibfield
  {journal} {\bibinfo  {journal} {PRX Quantum}\ }\textbf {\bibinfo {volume}
  {6}},\ \bibinfo {pages} {040367} (\bibinfo {year} {2025})}\BibitemShut
  {NoStop}%
\bibitem [{\citenamefont {Hebestreit}\ \emph
  {et~al.}(2018{\natexlab{b}})\citenamefont {Hebestreit}, \citenamefont
  {Frimmer}, \citenamefont {Reimann},\ and\ \citenamefont
  {Novotny}}]{hebestreit2018staticforces}%
  \BibitemOpen
  \bibfield  {author} {\bibinfo {author} {\bibfnamefont {E.}~\bibnamefont
  {Hebestreit}}, \bibinfo {author} {\bibfnamefont {M.}~\bibnamefont {Frimmer}},
  \bibinfo {author} {\bibfnamefont {R.}~\bibnamefont {Reimann}},\ and\ \bibinfo
  {author} {\bibfnamefont {L.}~\bibnamefont {Novotny}},\ }\bibfield  {title}
  {\bibinfo {title} {Sensing static forces with free-falling nanoparticles},\
  }\href {https://doi.org/10.1103/PhysRevLett.121.063602} {\bibfield  {journal}
  {\bibinfo  {journal} {Physical Review Letters}\ }\textbf {\bibinfo {volume}
  {121}},\ \bibinfo {pages} {063602} (\bibinfo {year}
  {2018}{\natexlab{b}})}\BibitemShut {NoStop}%
\bibitem [{\citenamefont {Bachor}\ and\ \citenamefont
  {Ralph}(2019)}]{bachor2019guide}%
  \BibitemOpen
  \bibfield  {author} {\bibinfo {author} {\bibfnamefont {H.-A.}\ \bibnamefont
  {Bachor}}\ and\ \bibinfo {author} {\bibfnamefont {T.~C.}\ \bibnamefont
  {Ralph}},\ }\href@noop {} {\emph {\bibinfo {title} {A Guide to Experiments in
  Quantum Optics}}},\ \bibinfo {edition} {3rd}\ ed.\ (\bibinfo  {publisher}
  {Wiley-VCH},\ \bibinfo {address} {Weinheim},\ \bibinfo {year}
  {2019})\BibitemShut {NoStop}%
\bibitem [{\citenamefont {Loudon}(2000)}]{loudon2000quantum}%
  \BibitemOpen
  \bibfield  {author} {\bibinfo {author} {\bibfnamefont {R.}~\bibnamefont
  {Loudon}},\ }\href@noop {} {\emph {\bibinfo {title} {The Quantum Theory of
  Light}}},\ \bibinfo {edition} {3rd}\ ed.\ (\bibinfo  {publisher} {Oxford
  University Press},\ \bibinfo {address} {Oxford},\ \bibinfo {year}
  {2000})\BibitemShut {NoStop}%
\bibitem [{\citenamefont {Arfken}\ \emph {et~al.}(2013)\citenamefont {Arfken},
  \citenamefont {Weber},\ and\ \citenamefont
  {Harris}}]{arfken2013mathematical}%
  \BibitemOpen
  \bibfield  {author} {\bibinfo {author} {\bibfnamefont {G.~B.}\ \bibnamefont
  {Arfken}}, \bibinfo {author} {\bibfnamefont {H.~J.}\ \bibnamefont {Weber}},\
  and\ \bibinfo {author} {\bibfnamefont {F.~E.}\ \bibnamefont {Harris}},\
  }\href@noop {} {\emph {\bibinfo {title} {Mathematical Methods for Physicists:
  A Comprehensive Guide}}},\ \bibinfo {edition} {7th}\ ed.\ (\bibinfo
  {publisher} {Academic Press},\ \bibinfo {address} {Waltham, MA},\ \bibinfo
  {year} {2013})\ Chap.~\bibinfo {chapter} {14},\ \bibinfo {note} {see Section
  14.1, Generating Function}\BibitemShut {NoStop}%
\bibitem [{\citenamefont {Bussgang}\ \emph {et~al.}(1952)\citenamefont
  {Bussgang} \emph {et~al.}}]{bussgang1952crosscorrelation}%
  \BibitemOpen
  \bibfield  {author} {\bibinfo {author} {\bibfnamefont {J.~J.}\ \bibnamefont
  {Bussgang}} \emph {et~al.},\ }\href@noop {} {\emph {\bibinfo {title}
  {Crosscorrelation functions of amplitude-distorted Gaussian signals}}},\
  \bibinfo {type} {Tech. Rep.}\ (\bibinfo  {institution} {Research Laboratory
  of Electronics, Massachusetts Institute of Technology},\ \bibinfo {year}
  {1952})\BibitemShut {NoStop}%
\bibitem [{\citenamefont {Gieseler}\ \emph {et~al.}(2021)\citenamefont
  {Gieseler}, \citenamefont {Gomez-Solano}, \citenamefont {Magazz{\`u}},
  \citenamefont {P{\'e}rez~Castillo}, \citenamefont {P{\'e}rez~Garc{\'i}a},
  \citenamefont {Gironella-Torrent}, \citenamefont {Viader-Godoy},
  \citenamefont {Ritort}, \citenamefont {Pesce}, \citenamefont {Arzola},
  \citenamefont {Volke-Sepulveda},\ and\ \citenamefont
  {Volpe}}]{gieseler2021optical}%
  \BibitemOpen
  \bibfield  {author} {\bibinfo {author} {\bibfnamefont {J.}~\bibnamefont
  {Gieseler}}, \bibinfo {author} {\bibfnamefont {J.~R.}\ \bibnamefont
  {Gomez-Solano}}, \bibinfo {author} {\bibfnamefont {A.}~\bibnamefont
  {Magazz{\`u}}}, \bibinfo {author} {\bibfnamefont {I.}~\bibnamefont
  {P{\'e}rez~Castillo}}, \bibinfo {author} {\bibfnamefont {L.}~\bibnamefont
  {P{\'e}rez~Garc{\'i}a}}, \bibinfo {author} {\bibfnamefont {M.}~\bibnamefont
  {Gironella-Torrent}}, \bibinfo {author} {\bibfnamefont {X.}~\bibnamefont
  {Viader-Godoy}}, \bibinfo {author} {\bibfnamefont {F.}~\bibnamefont
  {Ritort}}, \bibinfo {author} {\bibfnamefont {G.}~\bibnamefont {Pesce}},
  \bibinfo {author} {\bibfnamefont {A.~V.}\ \bibnamefont {Arzola}}, \bibinfo
  {author} {\bibfnamefont {K.}~\bibnamefont {Volke-Sepulveda}},\ and\ \bibinfo
  {author} {\bibfnamefont {G.}~\bibnamefont {Volpe}},\ }\bibfield  {title}
  {\bibinfo {title} {Optical tweezers: a comprehensive tutorial from
  calibration to applications},\ }\href {https://doi.org/10.1364/AOP.394888}
  {\bibfield  {journal} {\bibinfo  {journal} {Advances in Optics and
  Photonics}\ }\textbf {\bibinfo {volume} {13}},\ \bibinfo {pages} {74}
  (\bibinfo {year} {2021})}\BibitemShut {NoStop}%
\bibitem [{\citenamefont {Meade}(1983)}]{meade1983lockin}%
  \BibitemOpen
  \bibfield  {author} {\bibinfo {author} {\bibfnamefont {M.~L.}\ \bibnamefont
  {Meade}},\ }\href@noop {} {\emph {\bibinfo {title} {Lock-in Amplifiers:
  Principles and Applications}}},\ \bibinfo {series} {IEE Electrical
  Measurement Series}, Vol.~\bibinfo {volume} {1}\ (\bibinfo  {publisher}
  {Peter Peregrinus},\ \bibinfo {address} {London},\ \bibinfo {year}
  {1983})\BibitemShut {NoStop}%
\bibitem [{\citenamefont {Lyons}(2011)}]{lyons2011understanding}%
  \BibitemOpen
  \bibfield  {author} {\bibinfo {author} {\bibfnamefont {R.~G.}\ \bibnamefont
  {Lyons}},\ }\href@noop {} {\emph {\bibinfo {title} {Understanding Digital
  Signal Processing}}},\ \bibinfo {edition} {3rd}\ ed.\ (\bibinfo  {publisher}
  {Prentice Hall},\ \bibinfo {address} {Upper Saddle River, NJ},\ \bibinfo
  {year} {2011})\BibitemShut {NoStop}%
\bibitem [{\citenamefont {Tandeitnik}(2026)}]{tandeitnik2026levitationtoolbox}%
  \BibitemOpen
  \bibfield  {author} {\bibinfo {author} {\bibfnamefont {D.~R.}\ \bibnamefont
  {Tandeitnik}},\ }\href@noop {} {\bibinfo {title} {{LevitationToolbox}: A
  python toolbox for evaluating non-paraxial optical levitation and measurement
  backaction}},\ \bibinfo {howpublished}
  {\url{https://github.com/QuantumAdventures/LevitationToolbox}} (\bibinfo
  {year} {2026})\BibitemShut {NoStop}%
\bibitem [{\citenamefont {Millen}\ \emph {et~al.}(2014)\citenamefont {Millen},
  \citenamefont {Deesuwan}, \citenamefont {Barker},\ and\ \citenamefont
  {Anders}}]{millen2014nanoscale}%
  \BibitemOpen
  \bibfield  {author} {\bibinfo {author} {\bibfnamefont {J.}~\bibnamefont
  {Millen}}, \bibinfo {author} {\bibfnamefont {T.}~\bibnamefont {Deesuwan}},
  \bibinfo {author} {\bibfnamefont {P.}~\bibnamefont {Barker}},\ and\ \bibinfo
  {author} {\bibfnamefont {J.}~\bibnamefont {Anders}},\ }\bibfield  {title}
  {\bibinfo {title} {Nanoscale temperature measurements using non-equilibrium
  {B}rownian dynamics of a levitated nanosphere},\ }\href
  {https://doi.org/10.1038/nnano.2014.82} {\bibfield  {journal} {\bibinfo
  {journal} {Nature Nanotechnology}\ }\textbf {\bibinfo {volume} {9}},\
  \bibinfo {pages} {425} (\bibinfo {year} {2014})}\BibitemShut {NoStop}%
\bibitem [{\citenamefont {Ricci}\ \emph {et~al.}(2022)\citenamefont {Ricci},
  \citenamefont {Cuairan}, \citenamefont {Schell}, \citenamefont {Hebestreit},
  \citenamefont {Rica}, \citenamefont {Meyer},\ and\ \citenamefont
  {Quidant}}]{ricci2022chemical}%
  \BibitemOpen
  \bibfield  {author} {\bibinfo {author} {\bibfnamefont {F.}~\bibnamefont
  {Ricci}}, \bibinfo {author} {\bibfnamefont {M.~T.}\ \bibnamefont {Cuairan}},
  \bibinfo {author} {\bibfnamefont {A.~W.}\ \bibnamefont {Schell}}, \bibinfo
  {author} {\bibfnamefont {E.}~\bibnamefont {Hebestreit}}, \bibinfo {author}
  {\bibfnamefont {R.~A.}\ \bibnamefont {Rica}}, \bibinfo {author}
  {\bibfnamefont {N.}~\bibnamefont {Meyer}},\ and\ \bibinfo {author}
  {\bibfnamefont {R.}~\bibnamefont {Quidant}},\ }\bibfield  {title} {\bibinfo
  {title} {A chemical nanoreactor based on a levitated nanoparticle in
  vacuum},\ }\href {https://doi.org/10.1021/acsnano.2c01693} {\bibfield
  {journal} {\bibinfo  {journal} {ACS Nano}\ }\textbf {\bibinfo {volume}
  {16}},\ \bibinfo {pages} {8677} (\bibinfo {year} {2022})}\BibitemShut
  {NoStop}%
\bibitem [{\citenamefont {Hale}\ and\ \citenamefont
  {Querry}(1973)}]{hale1973optical}%
  \BibitemOpen
  \bibfield  {author} {\bibinfo {author} {\bibfnamefont {G.~M.}\ \bibnamefont
  {Hale}}\ and\ \bibinfo {author} {\bibfnamefont {M.~R.}\ \bibnamefont
  {Querry}},\ }\bibfield  {title} {\bibinfo {title} {Optical constants of water
  in the 200-nm to 200-$\mu$m wavelength region},\ }\href
  {https://doi.org/10.1364/AO.12.000555} {\bibfield  {journal} {\bibinfo
  {journal} {Applied Optics}\ }\textbf {\bibinfo {volume} {12}},\ \bibinfo
  {pages} {555} (\bibinfo {year} {1973})}\BibitemShut {NoStop}%
\bibitem [{\citenamefont {Zhang}\ \emph {et~al.}(2026)\citenamefont {Zhang},
  \citenamefont {Su}, \citenamefont {Hu}, \citenamefont {Yuan}, \citenamefont
  {Xiong}, \citenamefont {Han}, \citenamefont {Feng}, \citenamefont {Chen},
  \citenamefont {Kuang}, \citenamefont {Zhang}, \citenamefont {Han},
  \citenamefont {Xiao}, \citenamefont {Tan},\ and\ \citenamefont
  {Luo}}]{zhang2026mass}%
  \BibitemOpen
  \bibfield  {author} {\bibinfo {author} {\bibfnamefont {H.}~\bibnamefont
  {Zhang}}, \bibinfo {author} {\bibfnamefont {C.}~\bibnamefont {Su}}, \bibinfo
  {author} {\bibfnamefont {S.}~\bibnamefont {Hu}}, \bibinfo {author}
  {\bibfnamefont {J.}~\bibnamefont {Yuan}}, \bibinfo {author} {\bibfnamefont
  {W.}~\bibnamefont {Xiong}}, \bibinfo {author} {\bibfnamefont
  {D.}~\bibnamefont {Han}}, \bibinfo {author} {\bibfnamefont {H.}~\bibnamefont
  {Feng}}, \bibinfo {author} {\bibfnamefont {X.}~\bibnamefont {Chen}}, \bibinfo
  {author} {\bibfnamefont {T.}~\bibnamefont {Kuang}}, \bibinfo {author}
  {\bibfnamefont {Y.}~\bibnamefont {Zhang}}, \bibinfo {author} {\bibfnamefont
  {X.}~\bibnamefont {Han}}, \bibinfo {author} {\bibfnamefont {G.}~\bibnamefont
  {Xiao}}, \bibinfo {author} {\bibfnamefont {Z.}~\bibnamefont {Tan}},\ and\
  \bibinfo {author} {\bibfnamefont {H.}~\bibnamefont {Luo}},\ }\bibfield
  {title} {\bibinfo {title} {Mass measurement of an optically levitated
  nanoparticle via ratio fitting},\ }\href
  {https://doi.org/10.1016/j.measurement.2026.123541} {\bibfield  {journal}
  {\bibinfo  {journal} {Measurement}\ }\textbf {\bibinfo {volume} {292}},\
  \bibinfo {pages} {123541} (\bibinfo {year} {2026})}\BibitemShut {NoStop}%
\bibitem [{\citenamefont {Jain}\ \emph {et~al.}(2016)\citenamefont {Jain},
  \citenamefont {Gieseler}, \citenamefont {Moritz}, \citenamefont {Dellago},
  \citenamefont {Quidant},\ and\ \citenamefont {Novotny}}]{jain2016direct}%
  \BibitemOpen
  \bibfield  {author} {\bibinfo {author} {\bibfnamefont {V.}~\bibnamefont
  {Jain}}, \bibinfo {author} {\bibfnamefont {J.}~\bibnamefont {Gieseler}},
  \bibinfo {author} {\bibfnamefont {C.}~\bibnamefont {Moritz}}, \bibinfo
  {author} {\bibfnamefont {C.}~\bibnamefont {Dellago}}, \bibinfo {author}
  {\bibfnamefont {R.}~\bibnamefont {Quidant}},\ and\ \bibinfo {author}
  {\bibfnamefont {L.}~\bibnamefont {Novotny}},\ }\bibfield  {title} {\bibinfo
  {title} {Direct measurement of photon recoil from a levitated nanoparticle},\
  }\href {https://doi.org/10.1103/PhysRevLett.116.243601} {\bibfield  {journal}
  {\bibinfo  {journal} {Physical Review Letters}\ }\textbf {\bibinfo {volume}
  {116}},\ \bibinfo {pages} {243601} (\bibinfo {year} {2016})}\BibitemShut
  {NoStop}%
\bibitem [{\citenamefont {Richards}\ and\ \citenamefont
  {Wolf}(1959)}]{richards1959electromagnetic}%
  \BibitemOpen
  \bibfield  {author} {\bibinfo {author} {\bibfnamefont {B.}~\bibnamefont
  {Richards}}\ and\ \bibinfo {author} {\bibfnamefont {E.}~\bibnamefont
  {Wolf}},\ }\bibfield  {title} {\bibinfo {title} {Electromagnetic diffraction
  in optical systems {II}. {S}tructure of the image field in an aplanatic
  system},\ }\href {https://doi.org/10.1098/rspa.1959.0200} {\bibfield
  {journal} {\bibinfo  {journal} {Proceedings of the Royal Society of London.
  Series A}\ }\textbf {\bibinfo {volume} {253}},\ \bibinfo {pages} {358}
  (\bibinfo {year} {1959})}\BibitemShut {NoStop}%
\bibitem [{\citenamefont {Wolf}(1959)}]{wolf1959electromagnetic}%
  \BibitemOpen
  \bibfield  {author} {\bibinfo {author} {\bibfnamefont {E.}~\bibnamefont
  {Wolf}},\ }\bibfield  {title} {\bibinfo {title} {Electromagnetic diffraction
  in optical systems {I}. {A}n integral representation of the image field},\
  }\href {https://doi.org/10.1098/rspa.1959.0199} {\bibfield  {journal}
  {\bibinfo  {journal} {Proceedings of the Royal Society of London. Series A}\
  }\textbf {\bibinfo {volume} {253}},\ \bibinfo {pages} {349} (\bibinfo {year}
  {1959})}\BibitemShut {NoStop}%
\bibitem [{\citenamefont {Novotny}\ and\ \citenamefont
  {Hecht}(2012)}]{novotny2012principles}%
  \BibitemOpen
  \bibfield  {author} {\bibinfo {author} {\bibfnamefont {L.}~\bibnamefont
  {Novotny}}\ and\ \bibinfo {author} {\bibfnamefont {B.}~\bibnamefont
  {Hecht}},\ }\href@noop {} {\emph {\bibinfo {title} {Principles of
  Nano-Optics}}},\ \bibinfo {edition} {2nd}\ ed.\ (\bibinfo  {publisher}
  {Cambridge University Press},\ \bibinfo {address} {Cambridge},\ \bibinfo
  {year} {2012})\BibitemShut {NoStop}%
\bibitem [{\citenamefont {Hogenauer}(1981)}]{hogenauer1981}%
  \BibitemOpen
  \bibfield  {author} {\bibinfo {author} {\bibfnamefont {E.~B.}\ \bibnamefont
  {Hogenauer}},\ }\bibfield  {title} {\bibinfo {title} {An economical class of
  digital filters for decimation and interpolation},\ }\href
  {https://doi.org/10.1109/TASSP.1981.1163535} {\bibfield  {journal} {\bibinfo
  {journal} {IEEE Transactions on Acoustics, Speech, and Signal Processing}\
  }\textbf {\bibinfo {volume} {29}},\ \bibinfo {pages} {155} (\bibinfo {year}
  {1981})}\BibitemShut {NoStop}%
\bibitem [{\citenamefont {Volder}(1959)}]{volder1959}%
  \BibitemOpen
  \bibfield  {author} {\bibinfo {author} {\bibfnamefont {J.~E.}\ \bibnamefont
  {Volder}},\ }\bibfield  {title} {\bibinfo {title} {The {CORDIC} trigonometric
  computing technique},\ }\href {https://doi.org/10.1109/TEC.1959.5222693}
  {\bibfield  {journal} {\bibinfo  {journal} {IRE Transactions on Electronic
  Computers}\ }\textbf {\bibinfo {volume} {EC-8}},\ \bibinfo {pages} {330}
  (\bibinfo {year} {1959})}\BibitemShut {NoStop}%
\bibitem [{\citenamefont {Hebestreit}\ \emph
  {et~al.}(2018{\natexlab{c}})\citenamefont {Hebestreit}, \citenamefont
  {Reimann}, \citenamefont {Frimmer},\ and\ \citenamefont
  {Novotny}}]{hebestreit2018internal}%
  \BibitemOpen
  \bibfield  {author} {\bibinfo {author} {\bibfnamefont {E.}~\bibnamefont
  {Hebestreit}}, \bibinfo {author} {\bibfnamefont {R.}~\bibnamefont {Reimann}},
  \bibinfo {author} {\bibfnamefont {M.}~\bibnamefont {Frimmer}},\ and\ \bibinfo
  {author} {\bibfnamefont {L.}~\bibnamefont {Novotny}},\ }\bibfield  {title}
  {\bibinfo {title} {Measuring the internal temperature of a levitated
  nanoparticle in high vacuum},\ }\href
  {https://doi.org/10.1103/PhysRevA.97.043803} {\bibfield  {journal} {\bibinfo
  {journal} {Physical Review A}\ }\textbf {\bibinfo {volume} {97}},\ \bibinfo
  {pages} {043803} (\bibinfo {year} {2018}{\natexlab{c}})}\BibitemShut
  {NoStop}%
\bibitem [{\citenamefont {Ahn}\ \emph {et~al.}(2018)\citenamefont {Ahn},
  \citenamefont {Xu}, \citenamefont {Bang}, \citenamefont {Deng}, \citenamefont
  {Hoang}, \citenamefont {Han}, \citenamefont {Ma},\ and\ \citenamefont
  {Li}}]{ahn2018optically}%
  \BibitemOpen
  \bibfield  {author} {\bibinfo {author} {\bibfnamefont {J.}~\bibnamefont
  {Ahn}}, \bibinfo {author} {\bibfnamefont {Z.}~\bibnamefont {Xu}}, \bibinfo
  {author} {\bibfnamefont {J.}~\bibnamefont {Bang}}, \bibinfo {author}
  {\bibfnamefont {Y.-H.}\ \bibnamefont {Deng}}, \bibinfo {author}
  {\bibfnamefont {T.~M.}\ \bibnamefont {Hoang}}, \bibinfo {author}
  {\bibfnamefont {Q.}~\bibnamefont {Han}}, \bibinfo {author} {\bibfnamefont
  {R.-M.}\ \bibnamefont {Ma}},\ and\ \bibinfo {author} {\bibfnamefont
  {T.}~\bibnamefont {Li}},\ }\bibfield  {title} {\bibinfo {title} {Optically
  levitated nanodumbbell torsion balance and {GHz} nanomechanical rotor},\
  }\href {https://doi.org/10.1103/PhysRevLett.121.033603} {\bibfield  {journal}
  {\bibinfo  {journal} {Physical Review Letters}\ }\textbf {\bibinfo {volume}
  {121}},\ \bibinfo {pages} {033603} (\bibinfo {year} {2018})}\BibitemShut
  {NoStop}%
\end{thebibliography}%

\end{document}